\documentclass[usenatbib,usegraphicx]{mn2e}
\usepackage{amssymb}
\usepackage{multirow}

\title[The flexion formalism for extended sources]
{Shape, shear and flexion II -- Quantifying the flexion formalism for extended 
sources with the ray-bundle method\thanks{Research undertaken as part of the Commonwealth Cosmology Initiative (CCI: www.thecci.org), an international collaboration supported by the Australian Research Council}}

\author[C. J.~Fluke and P. D.~Lasky]
  {C.J.~Fluke$^1$\thanks{cfluke@swin.edu.au} and P.D.~Lasky$^{1,2}$ \\
  $^1$Centre for Astrophysics \& Supercomputing, 
	Swinburne University of Technology, 
	PO Box 218, Hawthorn, Victoria, 3122, Australia\\
  $^2$Theoretical Astrophysics, Eberhard Karls University of T\"ubingen, T\"ubingen 72076, Germany}

\date{Accepted, Jan 2011}

\pagerange{\pageref{firstpage}--\pageref{lastpage}} \pubyear{2011}

\def\LaTeX{L\kern-.36em\raise.3ex\hbox{a}\kern-.15em
    T\kern-.1667em\lower.7ex\hbox{E}\kern-.125emX}

\begin{document}

\label{firstpage}

\maketitle

\begin{abstract}
Flexion-based weak gravitational lensing analysis is proving to be a useful
adjunct to traditional shear-based techniques.  As flexion arises from gradients
across an image, analytic and numerical techniques are required to investigate
flexion predictions for extended image/source pairs.  Using the Schwarzschild
lens model, we demonstrate that the ray-bundle method for gravitational 
lensing can be used to accurately recover second flexion, and is consistent
with recovery of zero first flexion. 
Using lens plane to source plane 
bundle propagation, we find that second flexion can be recovered with an error 
no worse than $1\%$ for bundle radii smaller than $\Delta \theta = 0.01 
\theta_{\rm E}$ and lens plane impact pararameters greater than 
$\theta_{\rm E} + \Delta \theta$, where $\theta_{\rm E}$ is the angular
Einstein radius.  Using source 
plane to lens plane bundle propagation, we demonstrate the existence 
of a preferred flexion zone.  For images at radii closer to the lens than the 
inner boundary of this zone, indicative of the true strong lensing regime, 
the flexion formalism should be used with caution 
(errors greater than $5\%$ for extended image/source pairs).
We also define a shear zone boundary, beyond which image shapes are 
essentially indistinguishable from ellipses ($1\%$ error in ellipticity).
While suggestive that a traditional weak lensing analysis is satisfactory
beyond this boundary, a potentially detectable non-zero flexion signal remains.
\end{abstract}


\section{Introduction}
\label{sct:introduction}
Weak gravitational lensing provides one of the most direct probes of the matter 
distribution of the Universe as it is independent of both the dynamical 
state and the nature of the matter.  Building on the pioneering attempts
by \cite{valdes83} and \cite{tyson84} to measure coherent changes 
in the shapes of background galaxies due to a foreground lens population, weak lensing
techniques have now come of age.   In recent years, weak lensing
has successfully been applied in the cases of galaxy-galaxy lensing
\citep[e.g.][]{brainerd96,hudson98,fischer00,smith01,guzik02,hoekstra04,sheldon04,heymans06,mandelbaum06,parker07,johnston07,mandelbaum08}, lensing by clusters \citep[e.g.][]{smail97,wittman01,gray02,taylor04,gavazzi07,abate09,okabe09} and ``cosmic shear'' due to
large-scale structure \citep[e.g.][]{wittman00,bacon00,rhodes01,hoekstra02,refregier02,brown03,bacon03,heavens06,kitching07}.  For recent reviews of weak lensing
theory and applications, see \cite{schneider05} and \cite{hoekstra08}.

The conventional mathematical basis for weak lensing analysis assumes that a shear
field causes an additional ellipticity to the shape of a background source,
which can be calculated by measuring the moments of the images 
\citep{kaiser95}.   However, this approach does not account for higher-order 
shape distortions that occur when there are strong tidal fields across the 
image.  Recently, an additional lensing effect called flexion has been investigated 
as an extension to shear-based measurements \citep{goldberg02,goldberg05,bacon06}\footnote{see also \cite{irwin05,irwin06,irwin07} who independently derived equivalent higher-order gravitational lensing effects which they call {\it sextupole} lensing.}.  Flexion has two components denoted first and second flexion, which are a shift of the image centroid and a representation of the ``arciness'' of the image respectively.

One main issue associated with shear-based gravitational lensing is that galaxies are intrinsically elliptical in shape (ellipses at some inclination to the line-of-sight seen in projection),
so it is necessary to disentangle lens-induced shear from the intrinsic shape.  
Resolved galaxies, however, are not intrisincally flexed -- although  
systems undergoing a merger, or galaxies with substantial 
asymmetric sub-structure such as a large starforming region, may be mis-intepreted as flexion signals.
It has been suggested that flexion may provide a stronger constraint
on dark matter \citep{leonard07,bacon09,hawken09,leonard10}, galaxy cluster
mass models \citep{leonard09} and also on {\it delensing} gravitational wave signals \citep{shapiro10} than shear
on its own, notwithstanding the challenges in measuring flexion
(e.g. Okura, Umetsu \& Futamase 2007, 2008; Goldberg \& Leonard 2007;
Massey et al.  2007; Irwin \& Shmakova 2006; Schneider \& Er 2008).

In a previous paper we presented analytic flexion results for a range of popular 
mass density profiles: Schwarzschild lens, singular isothermal sphere (SIS), 
Navarro-Frenk-White (NFW) profile and S\'{e}rsic-like profiles 
(Lasky \& Fluke 2009; hereafter Paper I).  
Our analytic solutions present a flexion formalism where 
we consider a two-dimensional (2D) field in the lens plane, which allows for the
treatement of extended sources.  In this paper we extend our previous work by considering the following key question: over what range of image/source sizes is the flexion approach valid?  That is, if flexion appears 
as a gradient of shear across an image, which is implictly assumed to be zero 
for traditional weak-lensing analysis, how well can we recover flexion for 
extended sources?    

The approach we use is via the ray-bundle method 
introduced by Fluke, Webster \& Mortlock (1999).  Here, 
bundles of light rays with a known initial configuration are 
propagated through one or more lens planes, and the deflection of 
each light ray is determined using the gravitational lens equation, 
see equation (\ref{eqn:tle}) below.  The initial and 
final shapes of the bundles can now be used to obtain numerical estimates 
for convergence, shear, magnification, first and second flexion as 
a function of source size.  By starting with simple lens models, we 
use known analytic results to test the accuracy of our approach,
and hence assess its applicability to cases where
analytic flexion results do not exist (e.g. asymmetric mass profiles,
such as from $N$-body, dark matter halo simulations).

The paper is set out as follows: in Section \ref{sct:flexform}, we summarise 
the key results for the analytic flexion formalism. In Section \ref{sct:RBM}, 
we describe how the ray-bundle method can be used to obtain flexion along 
individual lines-of-sight. We present the results of our numerical testing,
and determine a range of image locations and bundle radii for 
which the flexion approximation is valid for a Schwarzschild lens model.  
We demonstrate the existence of a flexion zone, which defines a physical
region where the flexion formalism is valid for extended sources.
In Section \ref{sct:discussion} we present calculations
of the size of the flexion zone as a function of lens mass, lens and source
redshift.  
We present our conclusions and identify
future directions for this work in Section \ref{sct:conclusion}.

\section{The flexion formalism}
\label{sct:flexform}
In this section, we summarise relevant results of the flexion formalism -- for full
details, see Paper I and references therein.

\subsection{The gravitational lens equation}
\label{sct:theory}
The thin-lens gravitational lens equation relates the location 
of a background source and its image(s) due to an intervening mass 
distribution:
\begin{equation}
\eta_i = \frac{D_{S}}{D_{L}} \xi_i - D_{LS} \tilde{\alpha}_i.
\label{eqn:tle}
\end{equation}
For the simple case of a single lens at the origin of the lens plane, 
$\xi_i$ is the image location in the image plane, $\eta_i$ is 
the source location in the source plane, and $\tilde{\alpha}_i$ is the
deflection angle. Throughout this paper, indices $i,j,k=1,2$ signify 
the two-dimensional components, with summation assumed over repeated indices,
and tensors are implicity functions of the image position, 
$\xi_{i}$, unless otherwise indicated. 
The thin-lens approximation requires that the spatial extent of the 
lens distribution be much smaller than the angular diameter distances 
between the observer and lens, $D_{L}$, observer and source, $D_{S}$, 
and lens to source, $D_{LS}$.  A convenient scaling of the lens equation
into angular coordinates uses the substitutions $\beta_i = \eta_i/D_{S}$, 
$\theta_i = \xi_i/D_{L}$ and $\alpha_i = \frac{D_{LS}}{D_{S}}\tilde{\alpha}_i$, so 
we have the compact form:
\begin{equation}
\beta_i = \theta_i - \alpha_i.
\label{eqn:angle}
\end{equation}
We introduce a further scaling for the lens equation in section 3.4.

\subsection{The lens matrix}
As a mapping between the lens and source planes, equation (\ref{eqn:angle}) 
is often rewritten in terms of its Jacobian matrix:
\begin{equation}
A_{ij} = \frac{\partial \beta_i}{\partial \theta_j} = 
\left(
\begin{array}{cc}
1 - \kappa - \gamma_1 &  -\gamma_2 \\
-\gamma_2 & 1 - \kappa + \gamma_1
\end{array}
\right),
\end{equation}
where the matrix elements are identified with the convergence, $\kappa$, and 
two orthogonal components of shear, $\gamma_1$ and $\gamma_2$.  
The magnification 
along a given line-of-sight,
$\mu$, is related to the area distortion of the lens mapping:
\begin{equation}
\mu = 1/\det A_{ij} = 1/[(1-\kappa)^2 - \gamma^2],
\end{equation}
where the total shear is $\gamma = \sqrt{\gamma_1^2 + \gamma_2^2}$.

By setting the location of the light ray as the origin of coordinates in 
the lens and source planes, one finds for small changes in the position of the
light ray:
\begin{equation}
\label{eqn:delta}
\delta\beta_{i}=A_{ij}\delta\theta_{j}.
\end{equation}
The interpretation of this mapping is as follows: if the gradients of 
the shear and convergence do not change significantly across the image, 
then the first-order weak lensing effect is to produce an additional 
ellipticity in the shape of background sources.
While this approach is satisfactory for point sources, it breaks down for
extended sources.  This is seen even for the conceptually simplest lens model, 
the point mass or Schwarzschild lens, which produces increasingly ``arcy'' images 
for extended sources at small impact parameters to the lens.

In order to account for the effects of gradients in the shear and convergence,
a second-order Taylor expansion of the gravitational field is required: 
\begin{equation}
\delta\beta_i  \simeq A_{ij} \delta \theta_j + \frac{1}{2} D_{ijk} \delta \theta_j \delta \theta_k.
\label{eqn:Dijk}
\end{equation}
Following the approach of \citet{bacon06}, the tensor term, $D_{ijk}$, 
is associated with two additional lensing distortion terms 
\begin{equation}
D_{ijk} \equiv \frac{\partial A_{ij}}{\partial \theta_k} = {\cal F}_{ijk} + {\cal G}_{ijk}.
\end{equation}
By further defining ${\cal F}={\cal F}_{1}+i{\cal F}_{2}$ and ${\cal G}={\cal G}_{1}+i{\cal G}_{2}$, 
we note that the components of first (${\cal F}$) 
and second (${\cal G}$) flexion can be written: 
\begin{eqnarray}
\label{eqn:F1}
{\cal F}_1 & = &  -\frac{1}{2}(D_{111} + D_{122}),\\
\label{eqn:F2}
{\cal F}_2 & = &  -\frac{1}{2}(D_{211} + D_{222}),\\
\label{eqn:G1}
{\cal G}_1 & = &  -\frac{1}{2}(D_{111} - 3D_{122}),\\
\label{eqn:G2}
{\cal G}_2 & = &  -\frac{1}{2}(3 D_{211} - D_{222}).
\end{eqnarray}
The components of first and second flexion only require a 
subset of the tensor components, $D_{ijk}$, and do not depend on 
knowledge of $D_{112}$, $D_{121}$, $D_{212}$ or $D_{221}$.  The relevance 
of this is discussed in Section \ref{sct:quant}.

\subsection{Circularly symmetric lens models}
\label{sct:genlens}

For a circularly symmetric mass profile, the deflection angle is
\begin{equation} 
\tilde{\alpha}_i = \frac{4G}{c^2} \frac{M(\vert \xi \vert)}
{\vert \xi \vert^2} \xi_i,
\end{equation}
where $\vert \xi \vert = \sqrt{\xi_1^2 + \xi_2^2 }$, and the projected
mass, $M$, is
\begin{equation}
M(\vert \xi \vert) = 2\pi \int_0^{\vert \xi \vert} \Sigma(\xi')
\xi' {\rm d}\xi',\label{eqn:M}
\end{equation}
for surface density, $\Sigma(\vert \xi \vert)$. This enables us to
re-write equation (\ref{eqn:tle}) as
\begin{equation}
\eta_i = \xi_i \frac{D_S}{D_L} \left[1 - \frac{1}{\pi \Sigma_{\rm cr}}
\frac{M(\vert \xi \vert)}{\vert \xi \vert^2}\right], 
\end{equation}
with critical surface density
\begin{equation}
\Sigma_{\rm cr} = \frac{c^2}{4 \pi G} \frac{D_S}{D_L D_{LS}}.
\end{equation}
Defining the function
\begin{equation}
{\cal Q}(\zeta) := \pi \Sigma(\zeta) \zeta^2 - M(\zeta)
\end{equation}
we can now express the shear, convergence and first and second flexion with
their explicit dependence on physical coordinates and surface mass density
(see Paper I for full details):
\begin{eqnarray}
\kappa & = & \frac{\Sigma(\vert \xi \vert) }{\Sigma_{\rm cr}}, \label{eqn:sskap}\\
\gamma_1 & = & \frac{{\cal Q} (\vert \xi \vert)}{\pi \Sigma_{\rm cr} 
\vert \xi \vert^4} (\xi_1^2 - \xi_2^2),\label{eqn:ssg1} \\
\gamma_2 & = & \frac{2 {\cal Q}(\vert \xi \vert) }{\pi \Sigma_{\rm cr}
\vert \xi \vert^4} \xi_1 \xi_2, \label{eqn:ssg2}\\
{\cal F}_1 & = & \frac{D_{L}}{\Sigma_{\rm cr}} \frac{\partial \Sigma}
{\partial \xi_1}, \label{eqn:ssF1}\\
{\cal F}_2 & = & \frac{D_{L}}{\Sigma_{\rm cr}} \frac{\partial \Sigma}
{\partial \xi_2}, \label{eqn:ssF2}\\
{\cal G}_1 & = & \frac{D_L \xi_1 (\xi_1^2 - 3 \xi_2^2)}
{\pi \Sigma_{\rm cr} \vert \xi \vert^6}
\left[\pi \frac{{\rm d}\Sigma}{d\vert \xi \vert} \vert \xi \vert^3 - 4 {\cal Q}
(\vert \xi\vert)
\right],\label{eqn:ssG1} \\
{\cal G}_2 & = & \frac{D_L \xi_2 (3\xi_1^2 - \xi_2^2)}
{\pi \Sigma_{\rm cr} \vert \xi \vert^6}
\left[\pi \frac{{\rm d}\Sigma}{d\vert \xi \vert} \vert \xi \vert^3 - 4 {\cal Q}
(\vert \xi \vert)
\right].\label{eqn:ssG2}
\end{eqnarray}

\section{Flexion with the ray-bundle method}
\label{sct:RBM}

In this section, we describe how the ray-bundle method can be used as
a numerical means of estimating flexion along a given line-of-sight.  
We use the Schwarzschild lens model (hereafter, SL), as it provides us 
with simple analytic 
solutions for all of the relevant lensing properties.  However, since
the SL has zero first flexion everywhere except at the origin, we emphasise
the recovery of second flexion with the RBM, whilst demonstrating that
results remain consistent with zero first flexion.
We consider both backwards (lens plane to source plane) and forwards
(source plane to lens plane) ray-bundle propagation, as this separately
allows us to constrain the appropriate bundle radius to use and to quantify
the extent of the flexion zone.  

\subsection{Inverse ray-tracing}
While knowledge of an image location uniquely defines the source location,
in general, equation (\ref{eqn:tle}) is not easily invertible to give 
all image locations for a given source position.  While ``brute force'' 
solution methods can be used (e.g. Paczy\'{n}ski 1986), the main alternative
is to use inverse ray-tracing in its direct form, as 
was introduced by Kayser, Refsdal \& Stabell (1986) and Schneider 
\& Weiss (1986;1987), or in its hierachical tree-code form 
(Wambsganss 1990;1999).  Here, light rays are projected backwards from 
the observer, through the lens 
plane to the source plane, which is represented by a two-dimensional grid of 
source pixels.  The deflection of each light ray is calculated with equation
(\ref{eqn:tle}).  

While ray-tracing methods are extremely well-suited to studying statistical
lensing effects (e.g. the creation of magnification
maps for studying probabilities of high magnification events in quasar 
microlensing), they are less well-suited for studying lensing effects along
a given line-of-sight.  If it were possible to write down an explicit analytic 
form for the null geodesic equation of general relativity for an arbitrary 
mass (i.e. lens) distribution, the optical scalar equations (Sachs 1961; 
Dyer \& Roeder 1974) could be used to measure the changing shape 
of a (small) bundle of light rays as it propagates from the source to
the observer.  Unfortunately, specific solutions only exist for a limited
number of cases, such as light propogation in Swiss Cheese ``inhomogeneous'' 
cosmological models (Harper 1991), so an alternative approach is required.

\subsection{The ray-bundle method}
The ray-bundle method (RBM) was developed as an alternative 
to grid-based inverse ray-tracing for studying gravitational 
lensing along specified sightlines (Fluke, Webster \& Mortlock 1999).
An image shape is defined in the image plane (or first lens plane in 
the case of multiple lens planes), comprising a central light 
ray, which represents the ``null geodesic'', surrounded by a bundle 
of $N_{\rm ray}$ light rays.  The light rays in this bundle are 
traced backwards to the source plane, producing a distorted source shape.
As such, RBM provides a numerical analogy to the optical scalar solution,
while retaining properties of the inverse ray-tracing approach, such as 
the use of the thin lens equation for deflection calculations. 

In earlier work (Fluke, Webster \& Mortlock 2002), 
the RBM was used to obtain magnification 
probability distributions for dark matter-only cosmological 
models.  However, the RBM's emphasis on bundle shape means it is ideally 
suited for studying both linear and higher-order gravitational lensing 
phenomena (viz. shear and flexion) in detail along any given line-of-sight.  

\begin{figure}
\centering
\includegraphics[width=3.0in]{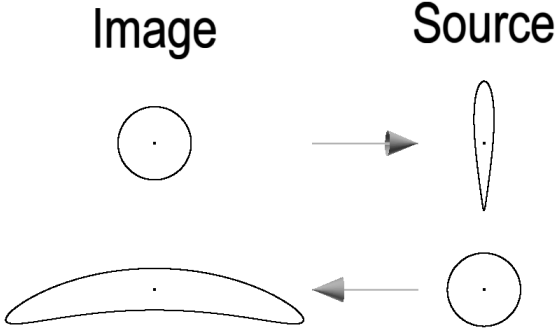}
\caption{\label{fig:shapes} Shapes in the image and source planes showing
non-linear lensing effects for extended image/source pairs.  
For the simplest case of a Schwarzschild lens, a circular image is mapped to 
a ``teardrop'' source (top row), and a circular source is mapped 
to an arc-like image (bottom row). }
\end{figure}

Moreover, the RBM gives a great deal of flexibility in choosing the size 
and shape of the initial bundle (i.e. the image), so that we can constrain
the length-scale over which the flexion approximation of
equation (\ref{eqn:Dijk}) is valid.  If the gradient of the shear and 
convergence across an image is negligible, i.e. we are in the regime of 
equation (\ref{eqn:delta}), 
which is the truly weak-field limit of gravitational 
lensing, then a circular source will appear as an elliptical image, 
and a circular image\footnote{although unlikely to occur naturally.} 
would be indicative of an intrinsically elliptical source.    
When tidal gradients {\em do} occur across an image, 
most readily due to the extended nature of image/source pairs, 
the relationship betwen image and source shapes is less obvious. 
For the SL, circular sources are mapped to 
``arc-like'' images, and circular images are due to ``teardrop'' shaped 
sources.  This correspondence is shown in Fig. \ref{fig:shapes}.

\subsection{Flexion with the ray-bundle method}
\label{sct:quant}
By analogy with equation (\ref{eqn:delta}), small changes in the position of 
a light ray in the image $(\delta \theta_1, \delta \theta_2)$ and 
source $(\delta \beta_1, \delta \beta_2)$ planes are related through the 
following equations, which are explicit expansions of equation (\ref{eqn:Dijk}):
\begin{eqnarray}
\delta \beta_1
\label{eqn:y1eq}
&=& A_{11} \delta \theta_1 + A_{12} \delta \theta_2 
+ \frac{1}{2} D_{111} \delta \theta_1^2 + \frac{1}{2} D_{122} \delta \theta_2^2 \\ \nonumber
&& + \frac{1}{2} \left[ D_{121} + D_{112} \right] \delta \theta_1 \delta \theta_2, \\
\delta \beta_2 
\label{eqn:y2eq}
&=& A_{21} \delta \theta_1 + A_{22} \delta \theta_2 
+ \frac{1}{2} D_{211} \delta \theta_1^2 + \frac{1}{2} D_{222} \delta \theta_2^2  \\ \nonumber
&& + \frac{1}{2} \left[ D_{212} + D_{221} \right] \delta \theta_1 \delta \theta_2.
\end{eqnarray}
By choosing the number of rays in each bundle, it is possible to obtain 
a complete solution for the unknown $A_{ij}$ and $D_{ijk}$ terms along 
an arbitrary line-of-sight.  Each light ray in the bundle experiences the 
tidal field around the central light ray, and so each pair of image
and source rays provides a unique solution to equations 
(\ref{eqn:y1eq}) and (\ref{eqn:y2eq}).  Since there are five unknowns we wish 
to solve for, we need only use six light rays in total per ray-bundle: 
a central light ray plus $N_{\rm ray} = 5$ rays defining the circumference.  
Writing these last equations in matrix form for each of the five $\delta \theta_{in}$ 
and $\delta \beta_{ik}$ bundle pairs ($k = 1 \dots 5$):
\begin{eqnarray}
\left(
\begin{array}{cccccc}
\delta \theta_{11} & \delta \theta_{21} & \delta \theta_{11} \delta \theta_{11} & \delta \theta_{11} \delta \theta_{21} & \delta \theta_{21} \delta \theta_{21} \\
\delta \theta_{12} & \delta \theta_{22} & \delta \theta_{12} \delta \theta_{12} & \delta \theta_{12} \delta \theta_{22} & \delta \theta_{22} \delta \theta_{22} \\
...\\
...\\
\delta \theta_{15} & \delta \theta_{25} & \delta \theta_{15} \delta \theta_{15} & \delta \theta_{15} \delta \theta_{25} &  \delta \theta_{25} \delta \theta_{25} \\
\end{array}
\right)
&&
\nonumber\\ 
\times \left(
\begin{array}{c}
A_{11} \\
A_{12} \\
\frac{1}{2} D_{111} \\
\frac{1}{2} [D_{112}+D_{121}] \\
\frac{1}{2} D_{122} \\
\end{array}
\right)
=
\left(
\begin{array}{c}
\delta \beta_{11}\\
\delta \beta_{12}\\
\delta \beta_{13}\\
\delta \beta_{14}\\
\delta \beta_{15}
\end{array}
\right),
&&
\label{eqn:simul1}
\end{eqnarray}

\begin{eqnarray}
\left(
\begin{array}{cccccc}
\delta \theta_{11} & \delta \theta_{21} & \delta \theta_{11} \delta \theta_{11} & \delta \theta_{11} \delta \theta_{21} & \delta \theta_{21} \delta \theta_{21} \\
\delta \theta_{12} & \delta \theta_{22} & \delta \theta_{12} \delta \theta_{12} & \delta \theta_{12} \delta \theta_{22} & \delta \theta_{22} \delta \theta_{22} \\
...\\
...\\
\delta \theta_{15} & \delta \theta_{25} & \delta \theta_{15} \delta \theta_{15} & \delta \theta_{15} \delta \theta_{25} &  \delta \theta_{25} \delta \theta_{25} \\
\end{array}
\right)
&&
\nonumber\\ 
\times \left(
\begin{array}{c}
A_{21} \\
A_{22} \\
\frac{1}{2} D_{211} \\
\frac{1}{2} [D_{212}+D_{221}] \\
\frac{1}{2} D_{222} \\
\end{array}
\right)
=
\left(
\begin{array}{c}
\delta \beta_{21}\\
\delta \beta_{22}\\
\delta \beta_{23}\\
\delta \beta_{24}\\
\delta \beta_{25}
\end{array}
\right).
&&
\label{eqn:simul2}
\end{eqnarray}
The unknown column vectors of the $A_{ij}$ and $D_{ijk}$ terms are the solutions 
to a set of simultaneous equations.  We note that there is a degeneracy in 
the co-effecients of the cross-terms, $\delta \theta_1 \delta \theta_2$, which 
means we cannot solve uniquely for $D_{112}$, $D_{121}$, $D_{212}$ 
and $D_{221}$.  Instead, these terms can be obtained from symmetries 
of $D_{ijk}$, viz. $D_{ijk} = D_{ikj}$ and $D_{12i} = D_{21i}$.
However, as we highlighted earlier, these four elements of the $D_{ijk}$ tensor 
do not contribute to first or second flexion [see equations 
(\ref{eqn:F1}--\ref{eqn:G2})],
and are instead associated with the twist and turn operators identified by
Bacon \& Sch\"{a}fer (2009).

\subsection{Image-to-source plane ray-bundles}
\label{sct:circimage}
Our first aim is to examine the (physical) scale in the image plane over which
the RBM can be used to recover flexion. 
Whereas we used angular coordinates for the lens equation and flexion
terms in sections 2.2 and 3.3, it is now more convenient to use a scaling 
relative to the size of the angular (point-mass) Einstein radius: 
\begin{equation}
\theta_{\rm E} = \sqrt{\frac{4 G M}{c^2}\frac{D_{LS}}{D_{S} D_{L}}}
\label{eqn:apmerad}
\end{equation}
in radians, for a lens with mass, $M$; $G$ and $c$ are the gravitational 
constant and speed of light respectively. 

In terms of scaled coordinates, $x_i = \theta_i/\theta_{\rm E}$
and $y_i = \beta_i/\theta_{\rm E}$, the two-dimensional lens equation 
for the SL model has solutions:
\begin{eqnarray}
\label{eqn:2dimage}
y_i & = & x_i \left[1 - \frac{1}{x_1^2 + x_2^2}\right]  \\
\label{eqn:2dsource}
x_i & = & \frac{y_i}{2} \left[1 \pm \sqrt{1 + \frac{4}{y_1^2 + y_2^2}}\,\right],
\hspace{0.5cm} y_1^2 + y_2^2 \neq 0,
\end{eqnarray}
so that the Einstein ring has a radius of 1 unit, and we can 
specify the bundle radius, $\Delta x$, as a fraction of the Einstein radius.
This scaling changes the Taylor expansions:
\begin{eqnarray}
\label{eqn:dy1}
\delta y_1 &=& A_{11} \delta x_1 + A_{12} \delta x_2 
+ \frac{\theta_{\rm E}}{2} \left[D_{111} \delta x_1^2 + D_{122} \delta x_2^2\right] \\ \nonumber
&&+ \frac{\theta_{\rm E}}{2} \left[D_{121} + D_{112}\right]
\delta x_1 \delta x_2 \\
\label{eqn:dy2}
\delta y_2 &=& A_{21} \delta x_1 + A_{22} \delta x_2 
+ \frac{\theta_{\rm E}}{2} \left[D_{211} \delta x_1^2 + D_{222} \delta x_2^2\right] \\ \nonumber
&&+ \frac{\theta_{\rm E}}{2} \left[D_{212} + D_{221}\right]
\delta x_1 \delta x_2
\end{eqnarray}
and the column vectors in equations (28) and (29) are now 
\begin{equation}
\label{eqn:colvec}
\left( 
\begin{array}{c}
A_{11}  \\
A_{12} \\
\frac{\theta_{\rm E}}{2} D_{111} \\
\frac{\theta_{\rm E}}{2} \left[D_{112} + D_{121}\right] \\
\frac{\theta_{\rm E}}{2} D_{122} \\
\end{array}
\right) \mbox{and}
\left( 
\begin{array}{c}
A_{21}  \\
A_{22} \\
\frac{\theta_{\rm E}}{2} D_{211} \\
\frac{\theta_{\rm E}}{2} \left[D_{212} + D_{221}\right] \\
\frac{\theta_{\rm E}}{2} D_{222} \\
\end{array}
\right) 
\end{equation}
once we substitute $\delta y_i = \delta \beta_i/\theta_{\rm E}$
and $\delta x_i = \delta \theta_i/\theta_{\rm E}$.
Note that the dimensionless nature of shear, through the $A_{ij}$ terms,
is maintained with this coordinate change: 
\begin{eqnarray}
\label{eqn:xgam1}
\gamma_1 &=& \frac{x_2^2 - x_1^2}{\left(x_1^2 + x_2^2\right)^2} \\
\label{eqn:xgam2}
\gamma_2 &=& \frac{-2 x_1 x_2}{\left(x_1^2 + x_2^2\right)^2}
\end{eqnarray}
as expected.

Next, we use the relationship $D_{ijk} = {\cal F}_{ijk} + {\cal G}_{ijk}$
(Bacon et al. 2006), which reduces to $D_{ijk} = G_{ijk}$ for the SL model, 
as all the $F_{ijk} = 0$. In terms of components, 
\begin{eqnarray}
D_{ij1} & = & {\cal G}_{ij1} = -\frac{1}{2} \left(
\begin{array}{ll}
{\cal G}_1 & {\cal G}_2\\
{\cal G}_2 & -{\cal G}_1
\end{array}
\right), \\
D_{ij2} & = & {\cal G}_{ij2} = +\frac{1}{2} \left(
\begin{array}{ll}
-{\cal G}_2 & {\cal G}_1\\
{\cal G}_2 & {\cal G}_2
\end{array}
\right)
\end{eqnarray}
and we use equations (\ref{eqn:ssG1}) and (\ref{eqn:ssG2}), 
to calculate the analytic solutions 
for the two components of second flexion for the SL model:
\begin{eqnarray}
\label{eqn:G1scale}
{\cal G}_1 & = & \frac{1}{\theta_{\rm E}}
\frac{4 x_1 \left(x_1^2 - 3 x_2^2\right)}
{\left(x_1^2 + x_2^2 \right)^3},\\
\label{eqn:G2scale}
{\cal G}_2 & = &  \frac{1}{\theta_{\rm E}}
\frac{4 x_2 \left(3 x_1^2 - x_2^2\right)}
{\left(x_1^2 + x_2^2 \right)^3}
\end{eqnarray}
Combining these last two equations implies 
\begin{equation}
\vert {\cal G} \vert = \frac{1}{\theta_{\rm E}} 
\frac{4}{\left( x_1^2 + x_2^2 \right)^{3/2}}.
\end{equation}
Fortuitously, the $\theta_{\rm E}^{-1}$ term 
in equations (\ref{eqn:G1scale}) and 
(\ref{eqn:G2scale}) effectively cancels the $\theta_{\rm E}$ coefficient
of the $D_{ijk}$ terms in the column vectors, equation (\ref{eqn:colvec}),
so we can obtain the scale independent second flexion with the RBM without
requiring a specific value  of $\theta_{\rm E}$.

Each ray-bundle is 
characterised by the radius, $\Delta x$, two-dimensional location of 
the central ray, $x_i$, and the number of bundle-rays, $N_{\rm ray}$.  As
shown in Section 3.3, we choose $N_{\rm ray} = 5$ in order to solve for
the unknown $A_{ij}$ and $D_{ijk}$ terms. 
For convenience in plotting and making comparisons with analytic 
results, we use polar coordinates, $(r, \phi)$, where 
$r = \sqrt{x_1^2 + x_2^2}$ is the radial impact parameter, and 
$\phi = \tan^{-1} (x_2/x_1)$ is the polar angle.  The analytic and 
RBM-recovered solutions are sampled on a grid with dimension $N_r 
\times N_\phi$ in polar space, which means that the sampling of solution
space is sparser with increasing $r$.  

For each $\Delta x$, we have to choose an appropriate range of $r$ values for sampling.
The minimum impact parameter is $r_{\rm min} = 1 + \Delta x(1 + \varepsilon)$,
where $\varepsilon \rightarrow 0$ is a small numerical offset (we use
$\varepsilon = 0.01$).  The offset avoids light rays in the image bundle 
overlapping the Einstein radius, thus ensuring that the $5 \times 5$ square matrix 
of $\delta x$ terms
is invertible.  While the lens
equation describes a mapping from the image plane backwards to 
the source plane, not all RBM images are permissable.  For example, it is
not possible to have a regular polygonal image straddling the Einstein
ring that corresponds to a single source shape.  However, the RBM does
work for images either completely outside the Einstein radius, or 
completely within it.\footnote{Images inside the Einstein radius are in 
the strong lensing regime, and hence outside the flexion zone that
we define in Section \ref{sct:circsource}}  For more complex lens models, where there
are non-degenerate caustics, the solution is less obvious.

The upper limit, $r_{\rm max}$, for the impact parameter is chosen to be
${\rm Max}\left(10, 10\Delta x\right)$, with the latter limit only relevant for bundle
radii greater than the Einstein radius.   In practice, the majority of bundle
radii we use were $\Delta x < 0.1$ (see Fig. 6). Our choice of $r_{\rm max} = 10$ 
was based on numerical tests, which included varying the number of radial
samples, $N_r$. Beyond this impact parameter, differences between the analytic
and RBM-recovered flexion values were consistent with numerical (i.e. precision) 
limits.  

For each bundle ray, $k = 1 \dots N_{\rm ray}$, we determine 
($\delta x_{1k}, \delta x_{2k})$ relative to the central ray, and either
apply  equations (\ref{eqn:dy1}) and (\ref{eqn:dy2}), assumed to be the exact analytic
solution, or utilise the RBM approach to deflect light rays with
the gravitational lens equation, to obtain $(\delta y_1, \delta y_2)$.
We then build the matrices of simultaneous equations for each ray-bundle
and use Gaussian elimination with back substitution to solve for the
$A_{ij}$ and $D_{ijk}$ terms, and hence ${\cal G}_i$ and ${\cal F}_i$.  

To test the matrix inversion code, we used the SL solutions of 
equations (\ref{eqn:xgam1}--\ref{eqn:xgam2}) and (\ref{eqn:G1scale}--\ref{eqn:G2scale}), 
as the analytic solutions for each $(r, \phi)$
sample.  Within the limits of numerical accuracy afforded by our
implementation, we find that we correctly recover the input
${\cal F}_i = 0$ and ${\cal G}_i$, independent of $N_r, N_\phi$ and $\Delta x$.
This demonstrates the accuracy and utility of our simultaneous equation-solving
code, under the proviso that there is no degeneracy in the mapping of
image ray-bundles to source ray-bundles. This assumption is appropriate in
the ``almost weak'' regime where flexion acts, and by ensuring that no
part of the source crosses the caustic point for the Schwarzschild lens.

Next, we use the scaled lens equation, equation (\ref{eqn:2dimage}), to deflect
the individual light rays in each bundle, so that we have pairs of image
and source bundle shapes, and build the matrices for each bundle.
Inverting the matrices, we obtain RBM-estimates for second flexion and use
recovered first flexion results as a consistency check on numerical effects (see
below).

\begin{figure}
\centering
\includegraphics[width=3.3in]{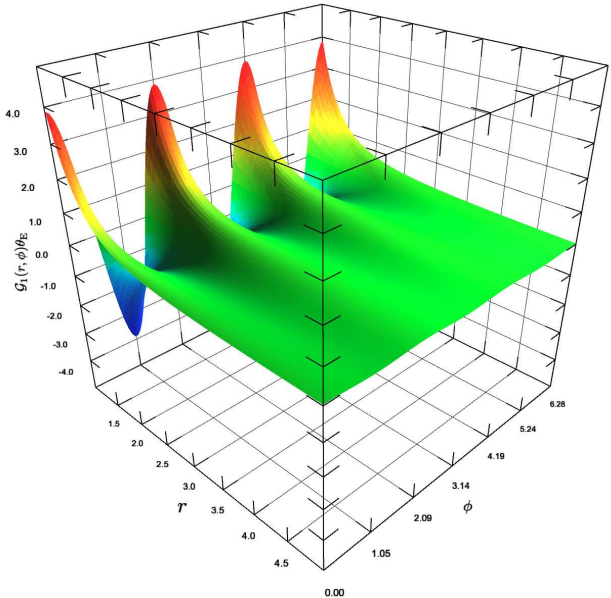}
\caption{\label{fig:G1surface} ${\cal G}_1(r, \phi) \theta_{\rm E}$ for
an image bundle radius $\Delta x = 0.01$, $r = \sqrt{x_1^2 + x_2^2}$ 
is the radial impact parameter in the range 
$1+\Delta x(1+0.01) \leq r \leq 5.0$,
$\phi = \tan^{-1}(x_2/x_1)$ and $x = \theta/\theta_{\rm E}$.  
The surface is sampled on an $100 \times 100$ polar grid. }
\end{figure}

In Fig. \ref{fig:G1surface}, we plot an indicative second flexion surface,
${\cal G}_1 (r, \phi) \theta_{\rm E}$; second flexion values must be
scaled by the angular Einstein radius (in radians) for comparison with a particular system.  
The spin-3 oscilliatory nature of second flexion is visible, with 
the amplitude increasing towards the origin.  The surface
is plotted over a limited range in impact parameter ($1.0101 \leq r \leq 5.0$), for
an image bundle with radius $\Delta x = 0.01$.  The surface is sampled over
a $N_r \times N_{\phi} = 100 \times 100$ polar grid.

We compare our RBM-based estimates of second flexion with the analytic values
using the mean-square error, $M_s$:
\begin{equation}
M_s = \frac{1}{N_1 N_2} \sum_{p=1}^{N_1} \sum_{q=1}^{N_2} \vert G(p,q)
- \hat{G}(p,q) \vert^2.
\end{equation}
Here, $G(p,q)$ is the grid-sampled analytic surface (one of ${\cal G}_i$ 
or $\vert {\cal G} \vert$), 
and $\hat{G}(p,q)$ is the grid-sampled, RBM estimate.
We define a peak signal-to-noise ratio as:
\begin{equation}
P_S = 20 \log_{10} \left[\frac{{\rm Max}(\hat{G})}{\sqrt{M_s}}\right],
\end{equation}
which provides a quantitative value for the equality of surfaces: $P_{S} \rightarrow
\infty$ if the surfaces are identical, and ${\rm Max}(\hat{G})$ is the maximum
value of the surface $\hat{G}(p,q)$. 
We find minimal dependence in the calculated $P_S$ on the
gridded surface resolution, with a variation $\Delta P_S = \pm 0.4$ when 
$P_S > 20$ for grids with dimensions $50 \leq N_r, N_\phi \leq 250$.

\begin{figure}
\centering
\includegraphics[width=3.0in,height=3.0in]{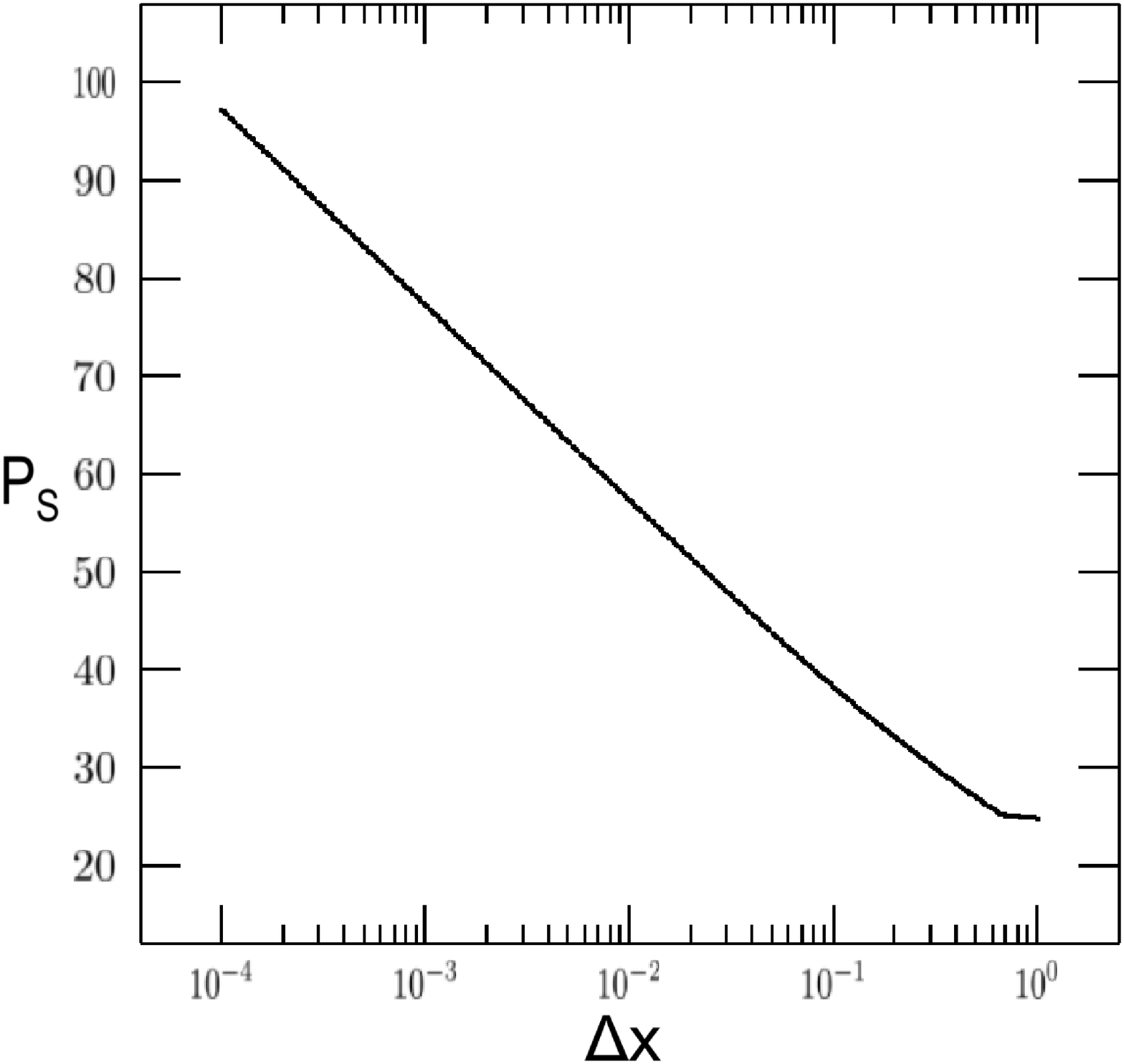}
\caption{\label{fig:G1power} Peak signal-to-noise ratio ($P_S$) versus image 
size, $\Delta x$, for RBM-recovery of ${\cal G}_1$. }
\end{figure}

We plot $P_S$ versus $\Delta x$ for RBM-recovery of ${\cal G}_1$
in Fig. \ref{fig:G1power}.  Results for ${\cal G}_2$ and
$\vert {\cal G} \vert$ are comparable with ${\cal G}_1$.
For the components of second flexion, the peak signal-to-noise
continues to grow as image size is reduced, and we approach the theoretical
point-source/point-image case. This result demonstrates that the RBM can indeed
be used to recover second flexion for the SL model, and gives us confidence
that this technique can also be applied for other, more complex lens models 
in a backwards ray-tracing mode (i.e. when the lens equation cannot be 
inverted to give image locations as a function of a source location).

Since the SL model has zero first flexion, we are not able to demonstrate
that the RBM is actually sensitive to the centroid shifts from a lens
model with non-zero first flexion components.  What we can show, however,
is that RBM is consistent with recovering zero first
flexion for the SL model. Using an approach similar to that outlined above
for ${\cal G}_i$ and $\vert {\cal G} \vert$, we determine estimates
for ${\cal F}_1$ from the matrix solutions 
as shown in Table \ref{tbl:tableF}.  The maximum (absolute) values of the
recovered ${\cal F}_1$ values are based on bundle radii in the range
$10^{-4} \leq \Delta x \leq 1$ and polar grid resolution $50 \leq N_r, N_\phi
\leq 500$. Results for ${\cal F}_2$ and $\vert {\cal F} \vert$ are comparable.

\begin{table}
\caption{Limits on recovery of first flexion, expected to be ${\cal F}_1 = 0$, 
for the Schwarzschild lens model using the RBM. $\Delta x$ is the 
image bundle radius in units of the Einstein radius.}
\label{tbl:tableF}
\begin{center}
\begin{tabular}{cc}
\hline
$\Delta x$ & Max$\left(\vert {\cal F}_1 \theta_{\rm E}\vert \right)$ \\
\hline
0.0001 & $1.1 \times 10^{-7}$ \\ 
0.001 & $2.2 \times 10^{-9}$ \\
0.01  & $1.9 \times 10^{-6}$ \\  
0.1 & $1.1 \times 10^{-3}$ \\  
1.0 & $3.1 \times 10^{-2}$ \\  
\hline
\end{tabular}
\end{center}
\end{table}

A slight increase in the maximum recovered first flexion is seen
at the lowest bundle radius: from $\sim 10^{-7}$ for $\Delta x = 10^{-4}$ 
to $\sim 10^{-9}$ for $\Delta x = 10^{-3}$, suggesting that we have 
reached a numerical limit (in our implementation).  We propose that 
$\Delta x = 10^{-3}$ is thus an appropriate lower bundle radius to 
maximise accuracy in using the RBM for an arbitrary lens configuration 
for recovering flexion. 

Due to our choice of $N_{\rm ray} =5$, there is a slight dependence on 
the orientation of the bundle with respect to the lens. Suppose we create
bundles that are evenly spaced in angle, $\phi$, around the lens, so 
that the bundle centres are:
\begin{equation}
{\mathbf I}_{c}(r, \phi) = [x_1, x_2] = 
\left[r \cos (\phi), r \sin (\phi)\right],
\end{equation} 
and the individual bundle rays are located at two-dimensional coordinates
\begin{equation}
{\mathbf I}_k (r, \phi) = 
\left[x_1 + \Delta x \cos(\psi_k + \psi_0),
x_2 + \Delta x \sin(\psi_k + \psi_0) \right],
\end{equation}
with $\psi_k = 2k\pi/N_{\rm ray}$ for $k=1 \dots N_{\rm ray}$, 
and $\psi_0$ is a phase term that is fixed per bundle.

\begin{figure}
\centering
\includegraphics[width=3.0in]{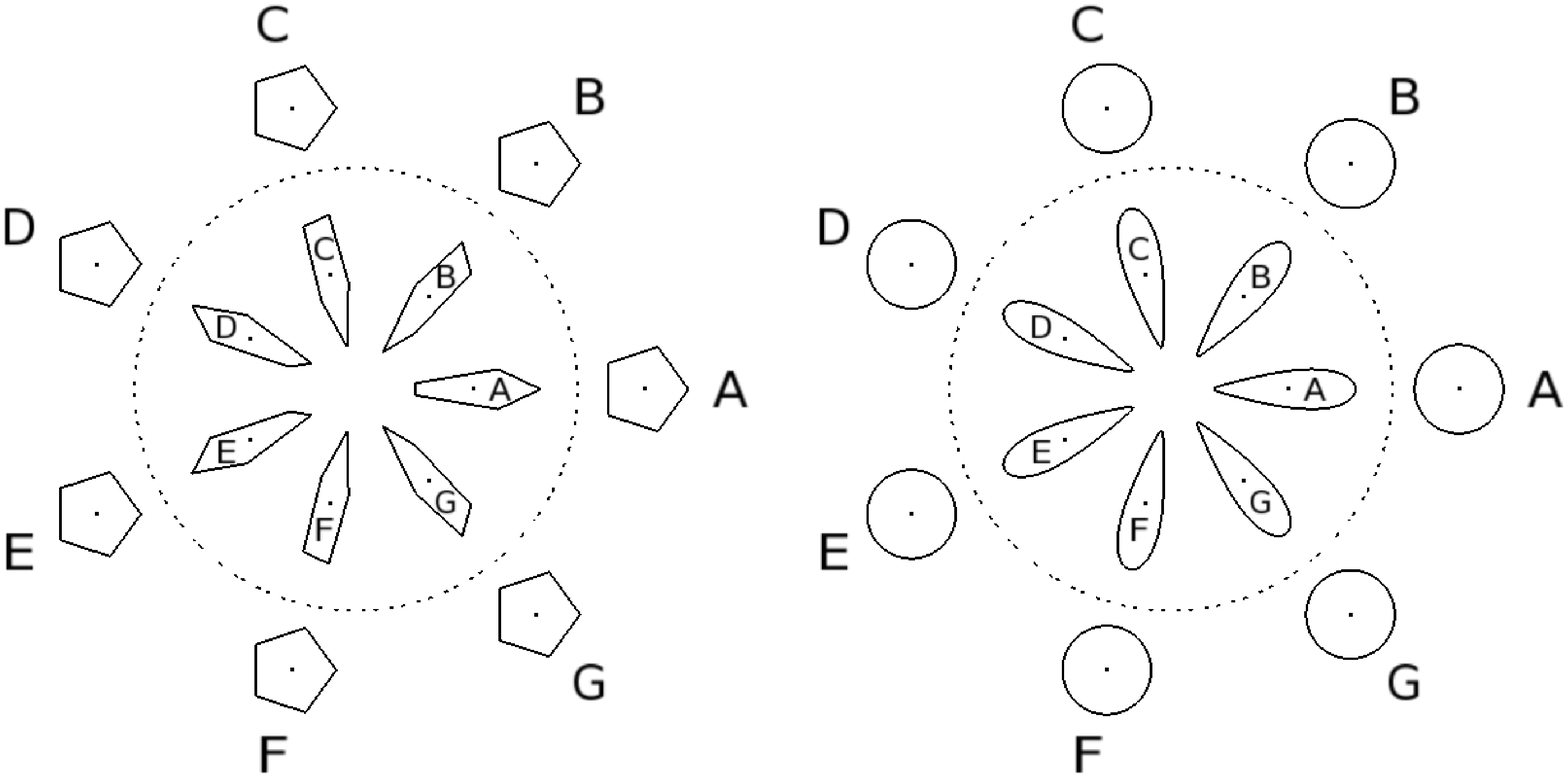}
\caption{\label{fig:penta} Pairs of sources and images represented as ray-bundles 
with $N_{\rm ray} = 5$ light rays around a central ray for circular images 
(thin lines) and circular source (thick lines).    The dotted line represents 
the Einstein radius for a Schwarzschild lens.  }
\end{figure}

If we set $\psi_0 = 0$ for all bundles, we get 
the situation shown in the left-hand panel of Fig. \ref{fig:penta}, where
the image (regular pentagons) and source bundles (irregular pentagons) are shown
with respect to the Einstein radius (dotted line). While the
bundles retain the same orientation in the image plane, independent of $\phi$,
the radial separation between the lens and the closest point of the bundle varies 
with $\phi$. This means that the tidal field experienced by each bundle
depends weakly on $\phi$.  For the case of image bundles that appear 
more circular (right-hand panel, $N_{\rm ray} = 100$) this effect is
much less pronounced, and we have the expected rotational symmetry.

\begin{figure*}
\centering
\includegraphics[width=3.2in]{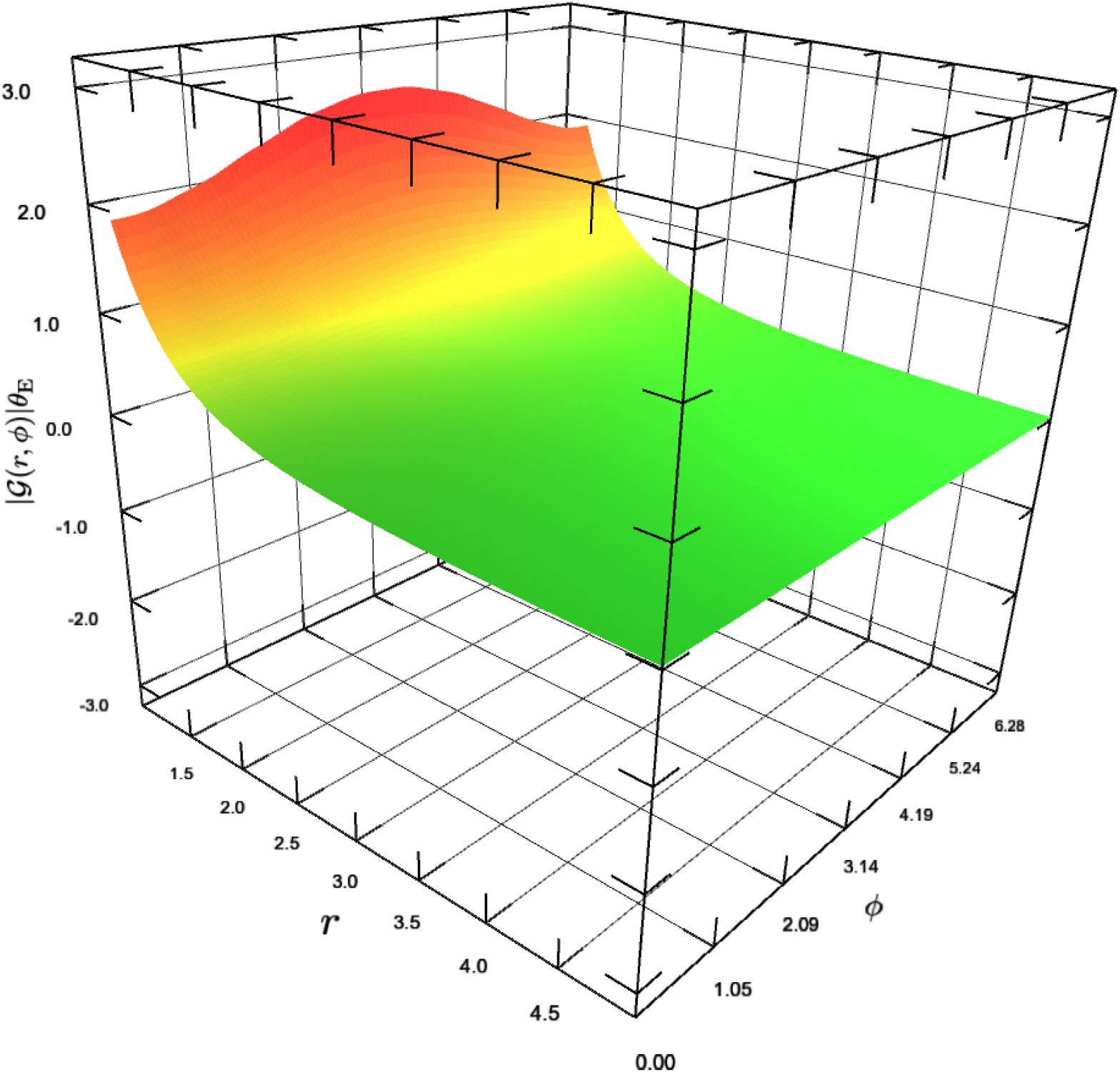}\hspace{-0.1cm}
\includegraphics[width=3.2in]{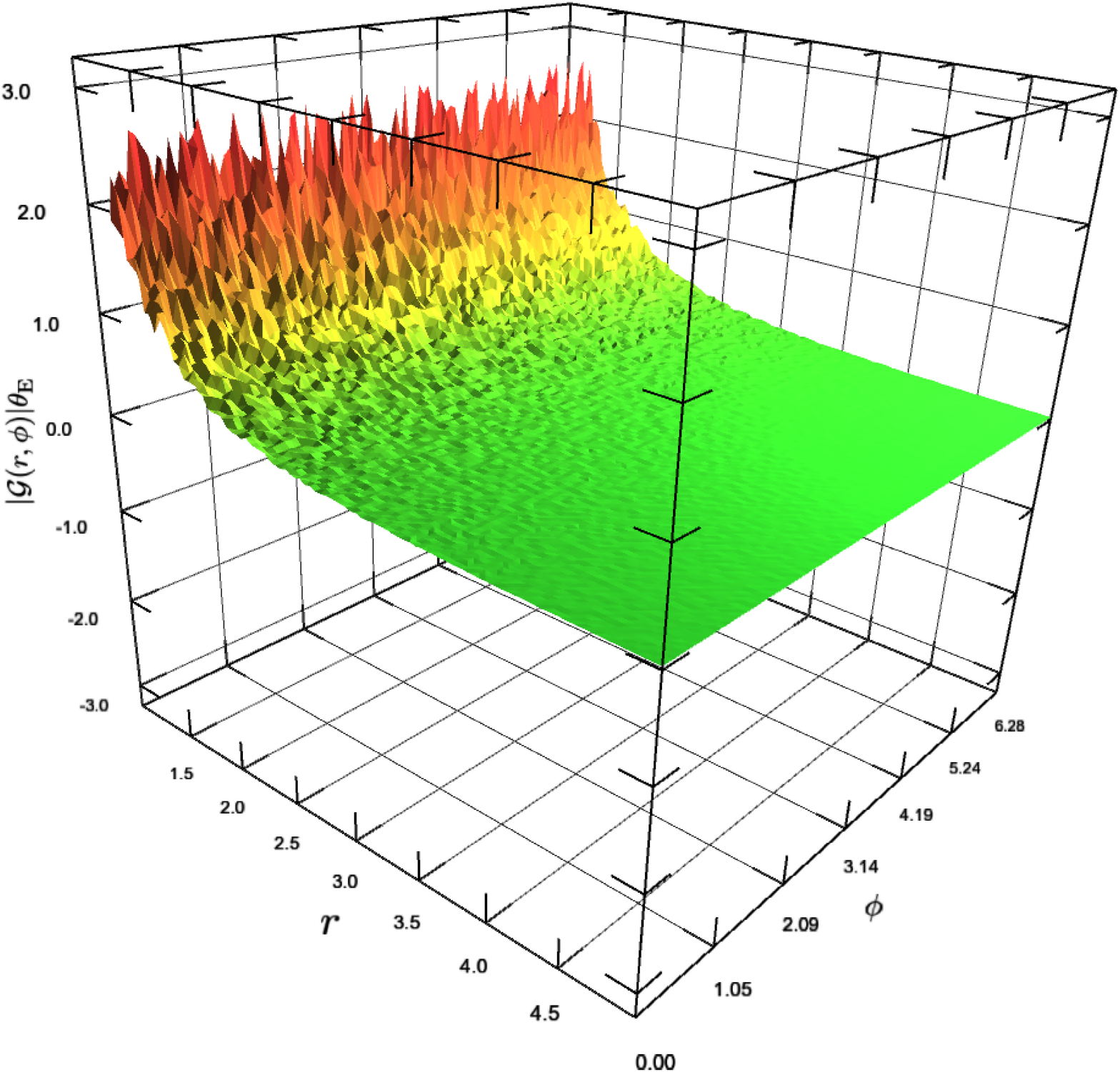}

\includegraphics[width=3.2in]{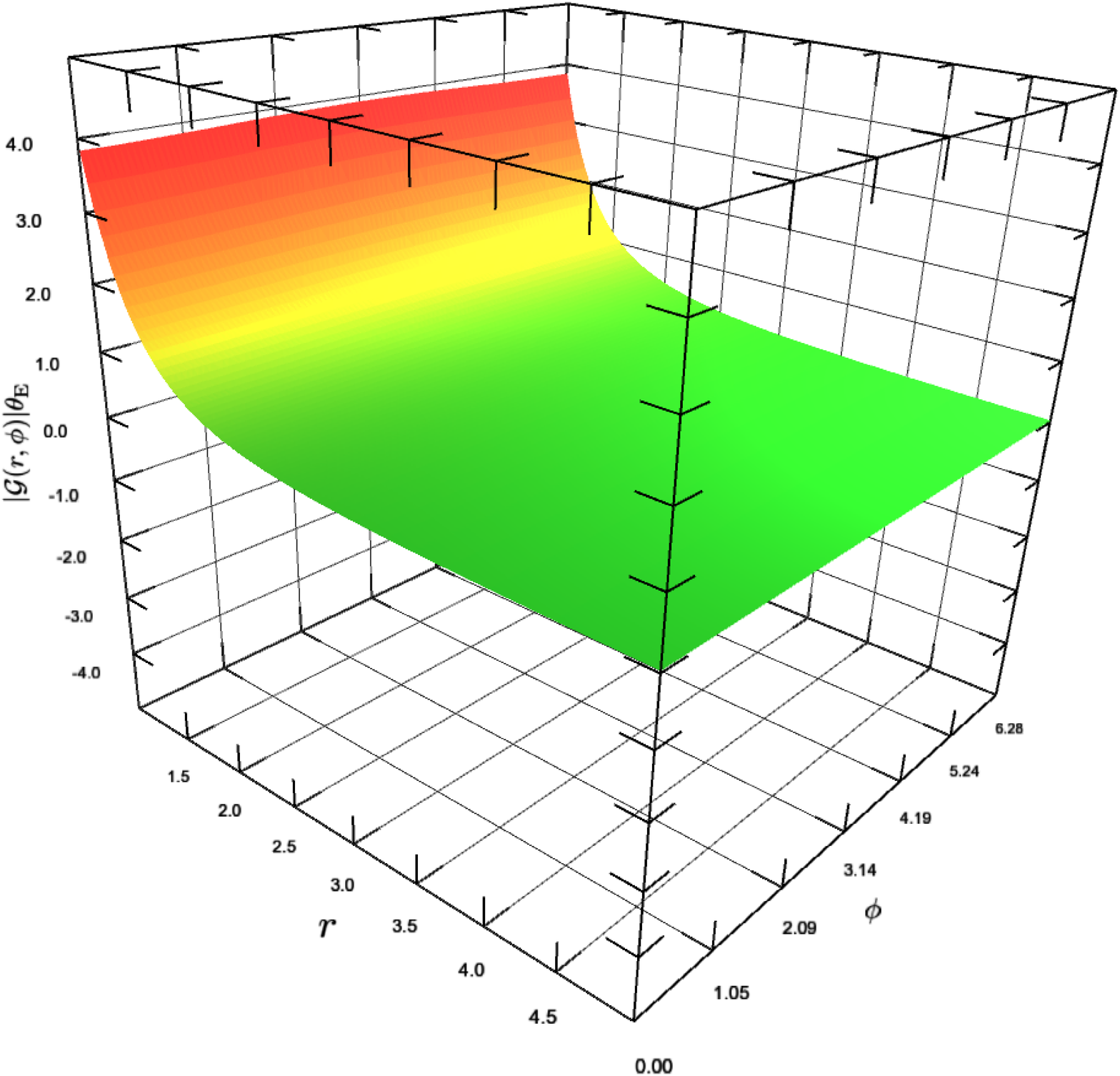}\hspace{-0.1cm}
\includegraphics[width=3.2in]{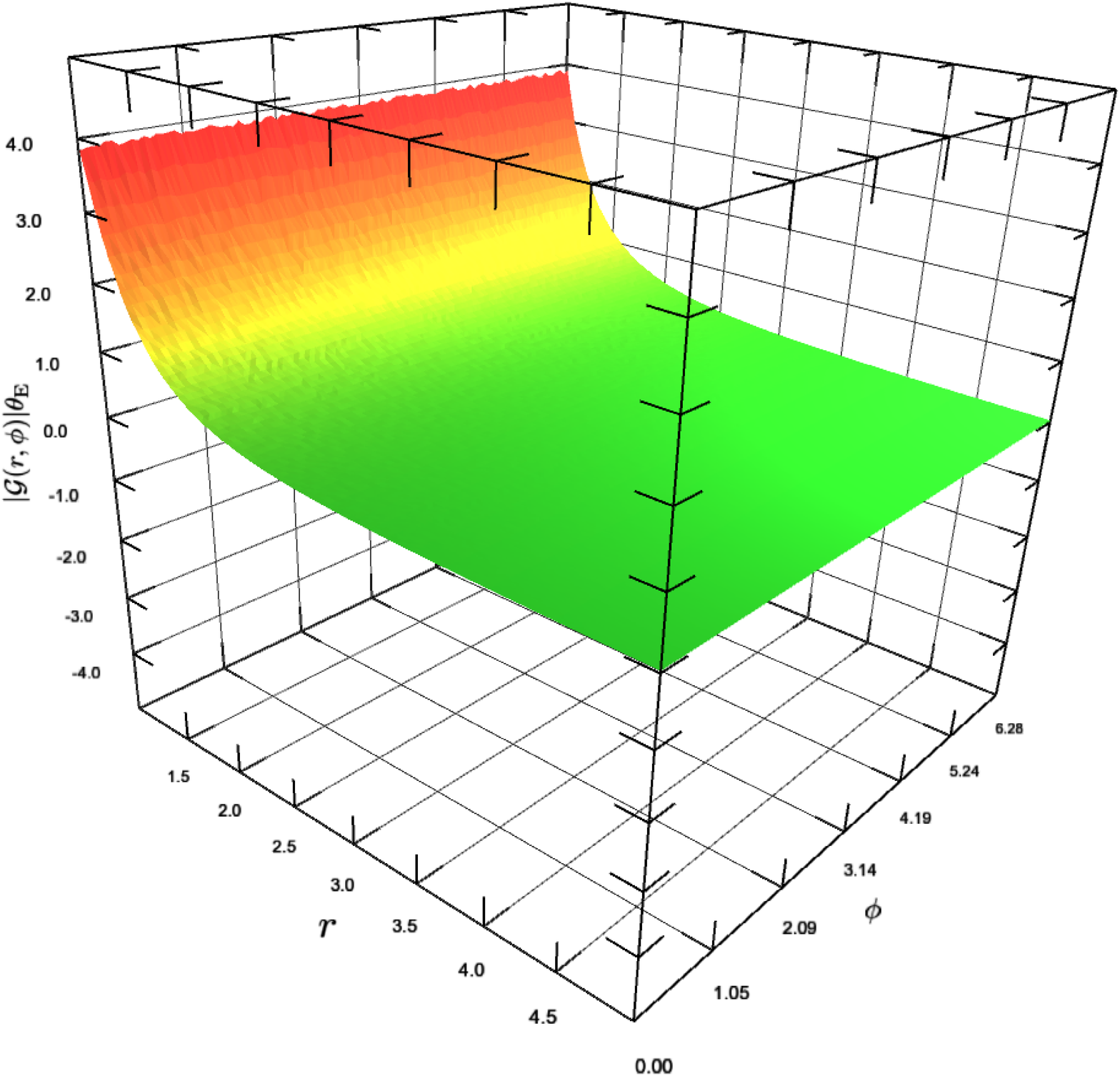}
\caption{\label{fig:varsurf} Sample $\vert {\cal G}(r, \phi)\vert \theta_{\rm E}$ 
surfaces obtained with the ray-bundle method, demonstrating the effect of the 
phase angle, $\psi_0$, for bundles in the image plane.  Each surface is calculated 
on a $100 \times 100$ polar grid, with impact parameters in the range $1 + \Delta x(1 + 0.01)
\leq r \leq 5$.
(Top left) Recovered with $\Delta x = 0.2$, bundle phase $\psi_0 = 0$.
(Top right) Recovered with $\Delta x = 0.2$, random $\psi_0$. 
(Bottom left) Recovered with $\Delta x = 0.01$, random $\psi_0$.
(Bottom right) Expected analytic $\vert {\cal G}(r,\phi) \vert\theta_{\rm E}$.  }
\end{figure*}

The consequence of this is demonstrated in Fig. \ref{fig:varsurf}, where
we plot $\vert \hat{{\cal G}}(r,\phi)\vert$ on a $100 \times 100$ polar grid
for $\Delta x = 0.2$ and $\psi_0 = 0$ (top left), $\Delta x = 0.2$
and uniform randomly selected $\psi_0$ (top right),  $\Delta x = 0.01$
and uniform randomly selected $\psi_0$ (bottom left) and analytic $\vert {\cal G} 
(r,\phi)\vert$ (bottom right).  A sinusoidal variation with $\phi$ is apparent when 
$\psi = 0$ for fixed $r$, and a scatter is introduced when $\psi_0$ 
is chosen uniformly randomly between $[0, 2\pi)$.   The magnitude of the
$\phi$ dependence is reduced by choosing a smaller $\Delta x$.

\begin{figure}

\begin{centering}
\includegraphics[width=3.2in]{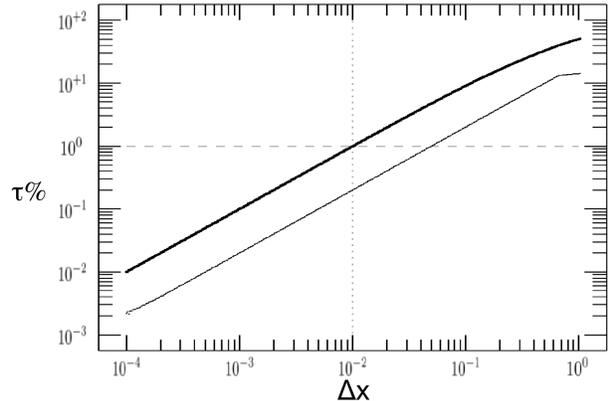}
\end{centering}

\caption{\label{fig:tau} Maximum (upper, thick line) and minimum 
(lower, thin line) 
percentage error, $\tau \%$ in RBM-recovered ${\cal G}(r)$ as
a function of image bundle radius, $\Delta x$.  }
\end{figure}

To quantify this effect we determine the minimum impact parameter, $r_{\tau}$,
at which $\left[G(r_{\tau}) - \hat{G}(r_{\tau})\right]/G(r) 
\leq \tau \%$ for a given image size, and 
$G({r_\tau}) = {\rm Max} \left[G(r,\phi)\right]$.  
In Fig. \ref{fig:tau} we plot the maximum (thick line) and minimum 
(thin line) percentage deviation between $\vert \hat{{\cal G}}(r) \vert$ 
and $\vert {\cal G}(r) \vert$ as a function of the image bundle radius.  
Samples were evaluated on a $250 \times 250$ polar grid.
The horizontal line is a nominal error of $1\%$, indicating that 
a sufficiently high level of 
accuracy occurs for $\Delta x = 0.01$ (vertical line).
Moreover, we find that for $\Delta x < 0.01$, the error 
in ${\cal G}_i$ is less than $1\%$ for all impact parameters $r > 1.01$, 
which suggests that we can use the RBM approach for impact parameters extending 
almost to the Einstein radius.  From this result, we infer that any variation we
see in the RBM-recovered $\hat{{\cal G}}_i$ for an arbitrary lens configuration
is evidence of a $\phi$-dependent second flexion, which differs from the predictions of a 
circularly-symmetric lens models.

\subsection{Source-to-image plane ray-bundles}
\label{sct:circsource}
For the SL we have the fortunate situation that there are 
analytic solutions for both $\theta \rightarrow  \beta$ (backwards) and 
$\beta \rightarrow \theta$ (forwards) versions of the
lens equation.  This gives us more 
flexibility in defining the image shape
in order to test the range over which the flexion formalism is appropriate.
In the previous section, we started with a regular image, which is deflected 
to a teardrop source.
Here, we use a regular polygonal source, and obtain the corresponding image shape. 
Propagating this modified (i.e. flexed) image backwards as a ray-bundle, 
we hope to recover the input source shape.

We note that an image can cover 
a small portion of the image plane, and hence the tidal fields can be
weak over the extent of the image.  For a strongly flexed (``banana-shaped'')
image, more of the image plane is covered by the image, 
and hence tidal effects are more prevalent.  For an arbitrary lens 
model or configuration, where we do not have analytic solutions 
for $\beta \rightarrow \theta$, we must work in the ``regular polygonal image 
maps to unknown source shape'' regime, so it is critical to understand 
the limits of our technique.

\begin{figure*}
\centering
\includegraphics[width=3.0in]{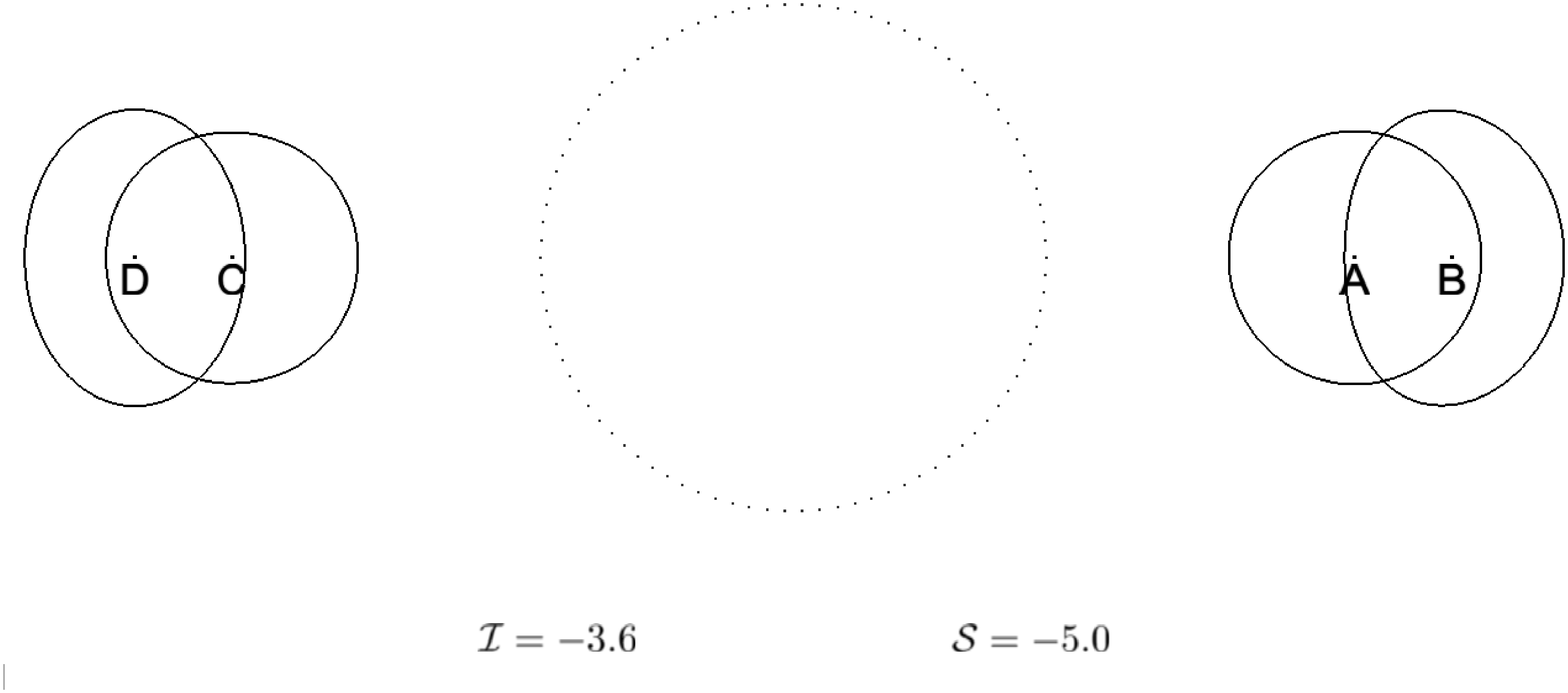}
\includegraphics[width=3.0in]{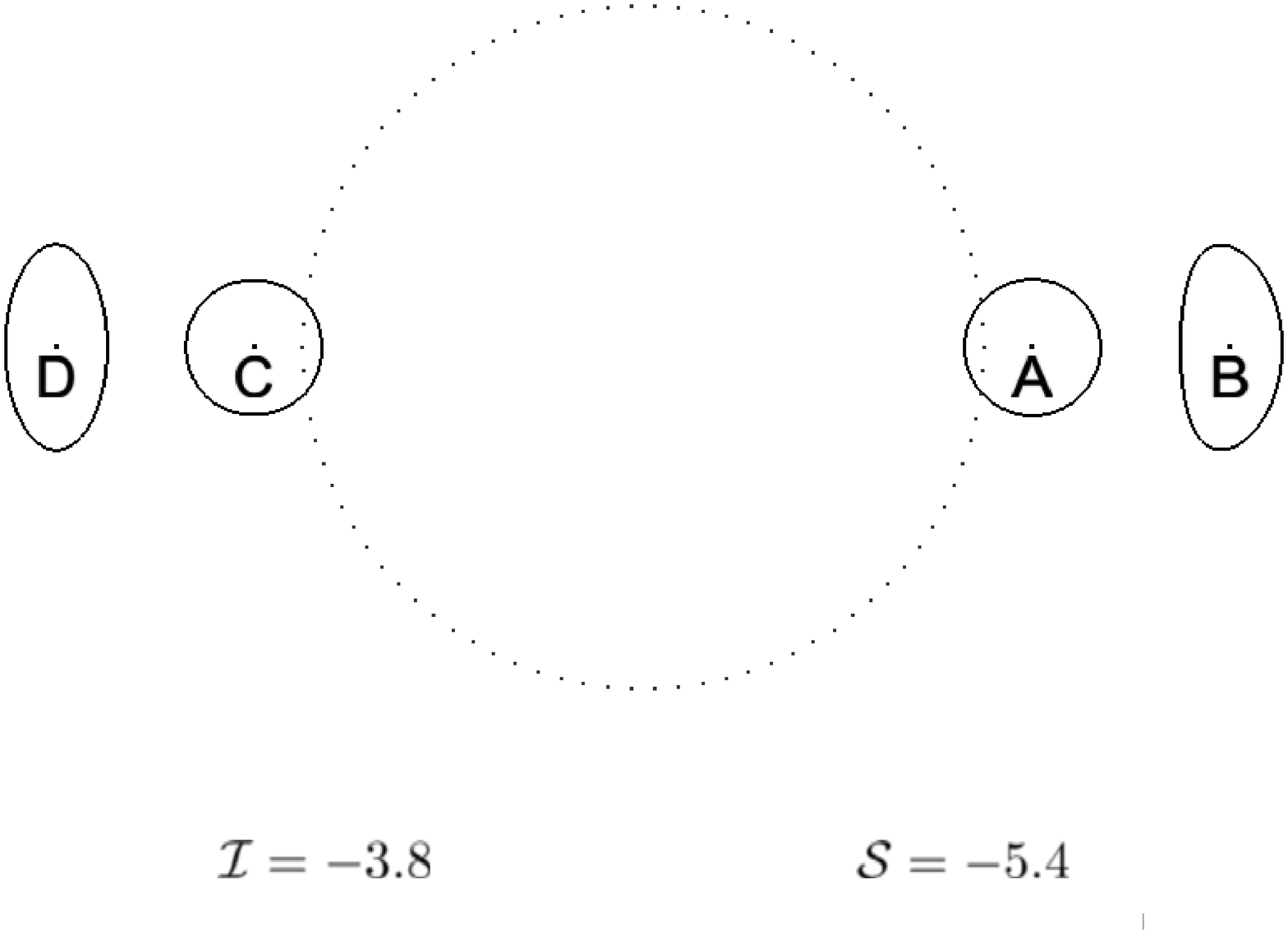}

\includegraphics[width=3.0in]{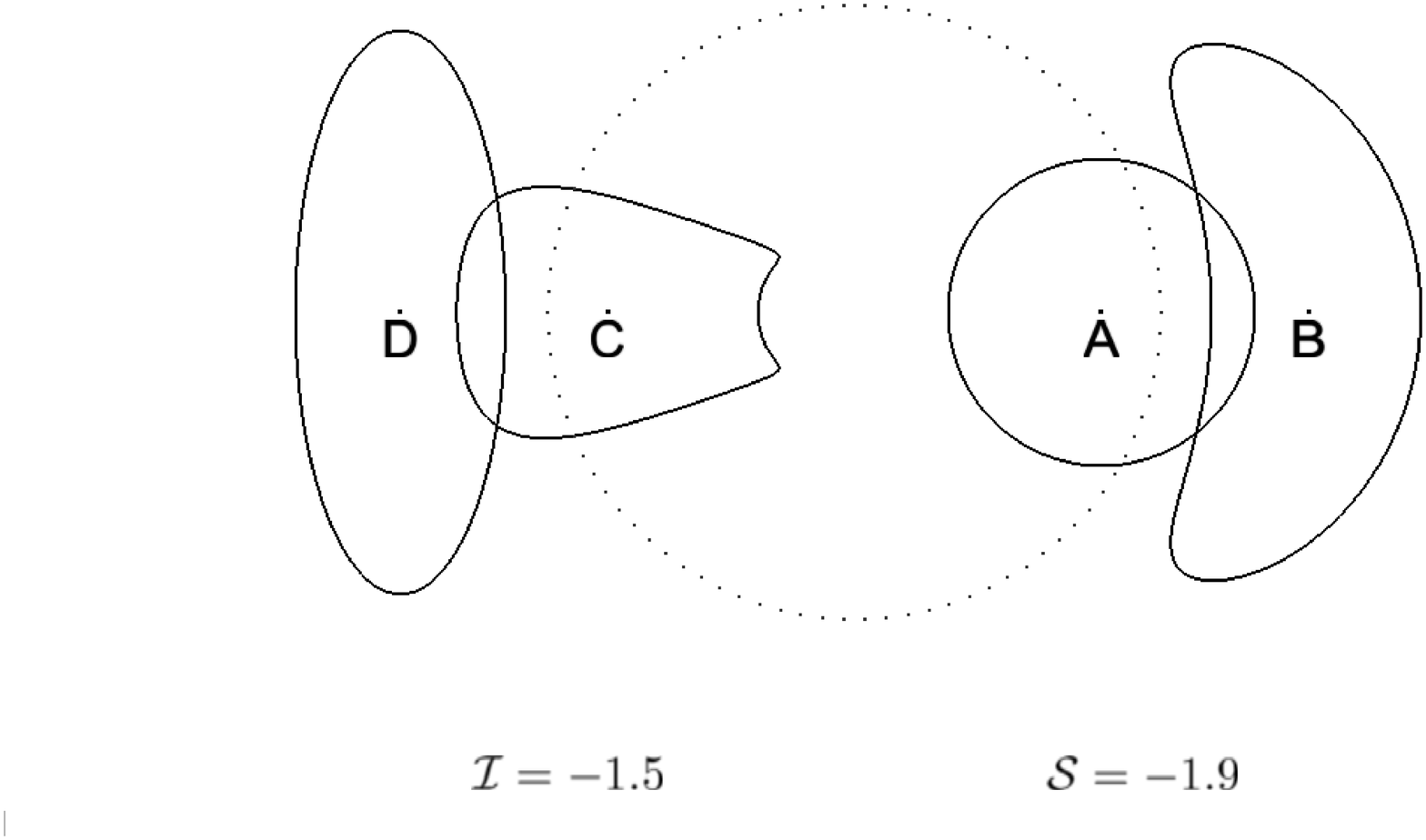}
\includegraphics[width=3.0in]{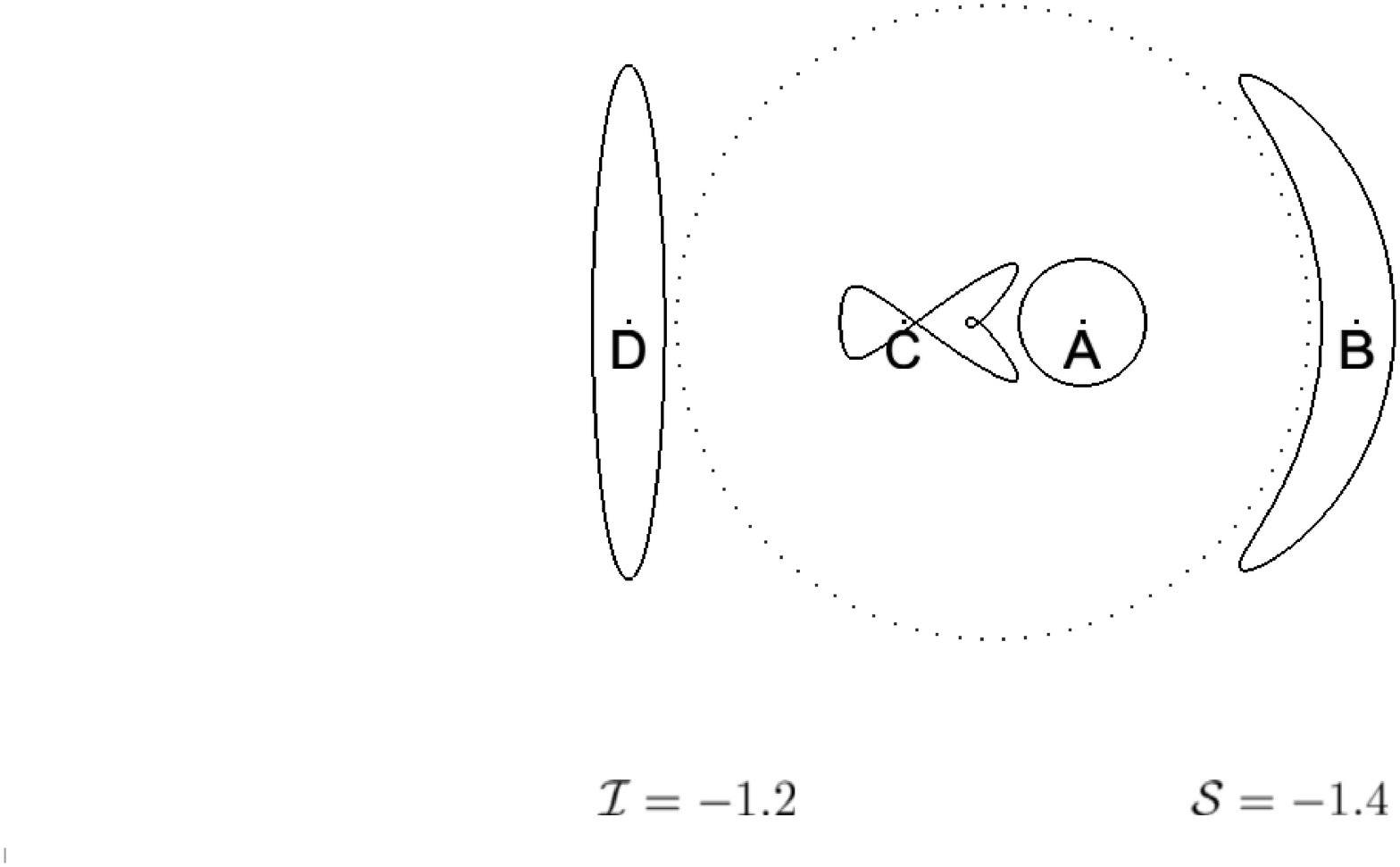}

\includegraphics[width=3.0in]{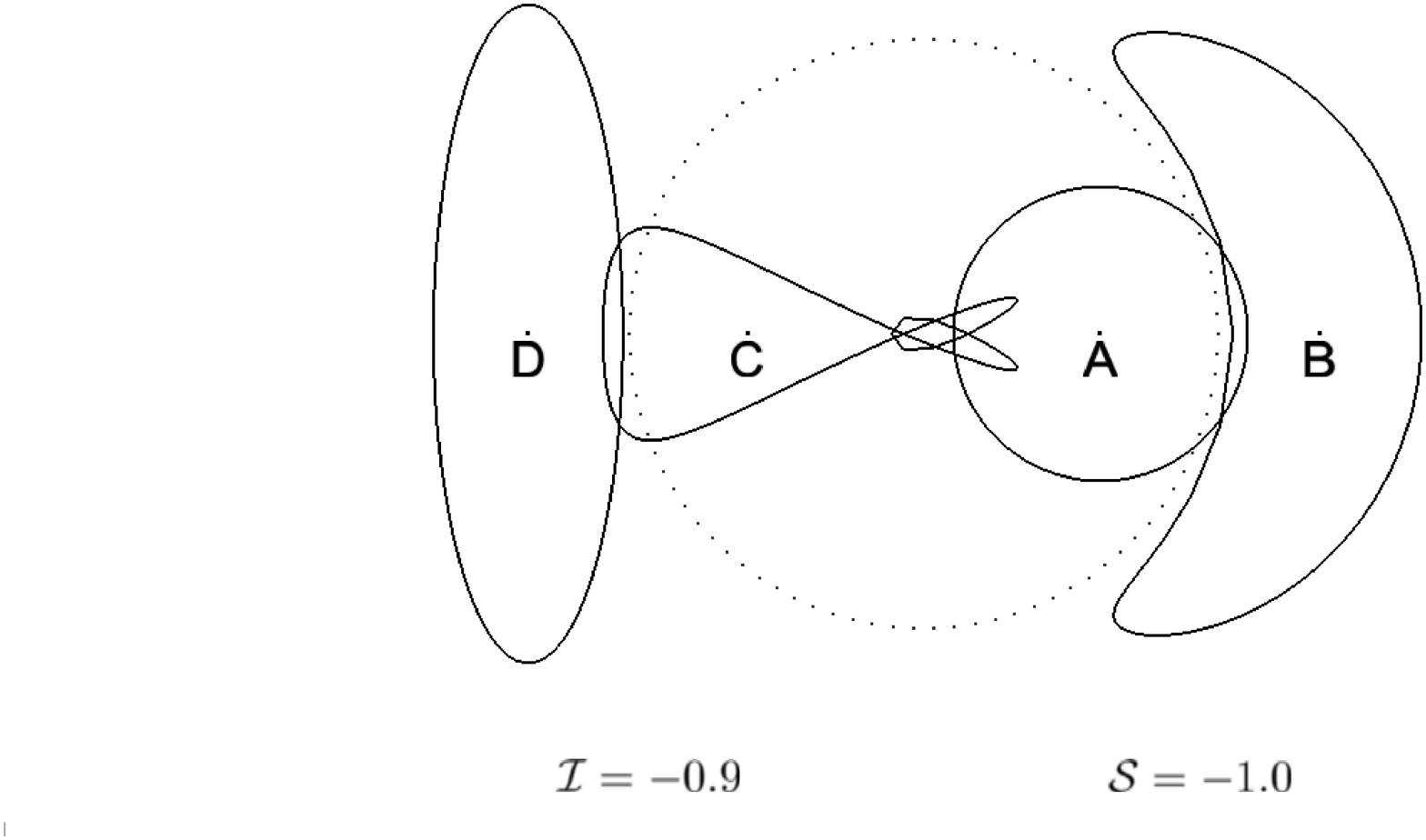}
\includegraphics[width=3.0in]{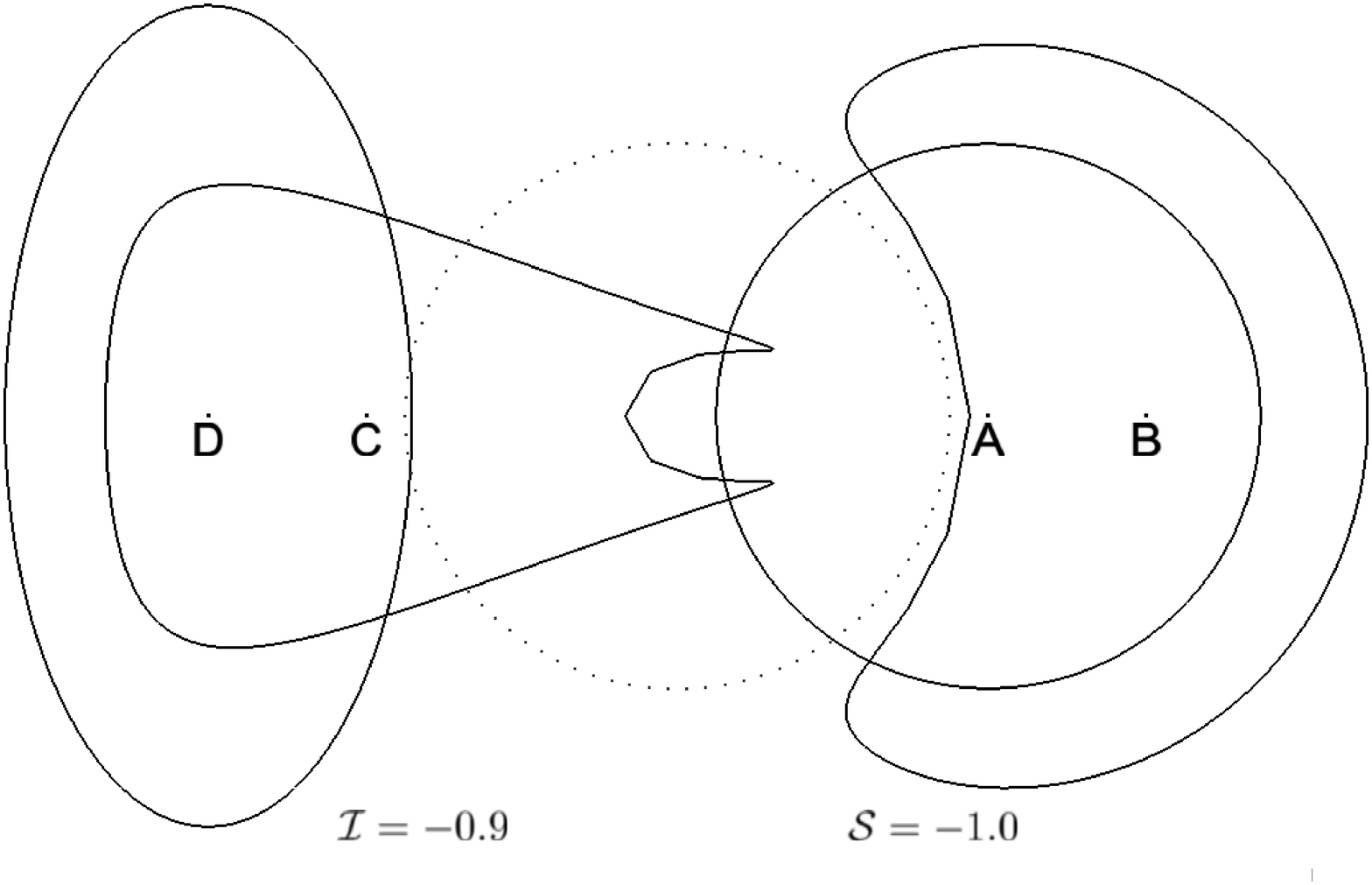}

\caption{Sample plots showing the limitations of the flexion formalism
for extended sources using the Schwarzschild lens model, with respect
to the Einstein radius (dotted circles).
In each case, a circular source (A) is mapped as a ray-bundle 
to a flexed image (B) using the lens equation (i.e. reverse RBM).  
This flexed image is mapped back to the source plane (C) using equation 
(\ref{eqn:Dijk}). The elliptical, shear-only version of the image (D), 
obtained by applying equation (\ref{eqn:delta}) to the circular source, is 
shown for comparison.  Note that source C and image D are presented as mirror
images of their actual positions for clarity in this figure. Analytic 
expressions for flexion of extended sources are valid in cases where 
source shapes A and C are comparable; shear-only weak lensing analysis
is satisfactory when image shapes B and D are comparable. 
The level of discrepancy between true and recovered source shapes 
increases for larger image radii and small impact parameters. Each 
bundle comprises 100 rays.  
Source bundle radii are: $\Delta y = 0.5$ (left column), 
0.2 (top right and middle right), and 1.0 (bottom right). ${\cal I}$ 
and ${\cal S}$ values are noted for each scenario, as defined in equations 
(\ref{eqn:Ical}) and (\ref{eqn:Scal}). }
\label{fig:faceplot}
\end{figure*}

Fig.~\ref{fig:faceplot} shows six sample configurations. 
In each case, a circular source (A) is mapped to a flexed image (B) using the 
inverse solution, equation (\ref{eqn:2dsource}), of the lens equation for the SL.
Next, equation (\ref{eqn:Dijk}) maps the 
differential image ray coordinates, $\delta \theta_i$, back to the source
plane to produce a new source profile (C). 
Finally, equation
(\ref{eqn:delta}) is used to create an elliptical, shear-only image shape (D) from
the circular source.  The major and minor axis lengths of image D are 
comparable to the equivalent (distorted) axes for the flexed image (C).  Source
C and image D are presented as mirror images of their true positions for clarity
in the figure.   We have used $N_{\rm ray} = 100$ for the bundles in 
this figure in order to show the true shapes,  
hence we can neglect the $\phi$-dependence (see Section \ref{sct:circimage}). 
While the Scwharzschild lens
model produces two images for each source position, we only consider
the more highly magnified image outside of the Einstein radius.
Significant variation between  the source shapes, A and C, is an indication that 
we are in a regime where the second order Taylor series expression for flexion is not valid 
for extended sources - this is the true strong lensing regime where images are 
arcs rather than arclets.

We examine the level of agreement between the Taylor expansion,
equation (\ref{eqn:Dijk}), and the forwards RBM solution 
by comparing the relative locations of each image or source ray in the bundle.  
We introduce two quantities:
\begin{equation}
{\cal I} = \log_{\rm 10} \sigma_I^2 = \log_{10} \left(\frac{1}{N_{\rm ray}} \sum_{n=1}^{N_{\rm ray}} \vert \mathbf{I}_{B,n} - \mathbf{I}_{D,n}' \vert^2 \right),
\label{eqn:Ical}
\end{equation}
where the two-dimensional vectors, $\mathbf{I}_{B,n}$ and $\mathbf{I}_{D,n}'$, 
are 
the $N_{\rm ray}$ light rays in the actual (B) and elliptical (D) images, and
similarly
\begin{equation}
{\cal S} = \log_{\rm 10} \sigma_S^2 = \log_{10} \left( \frac{1}{N_{\rm ray}} 
\sum_{n=1}^{N_{\rm ray}} \vert \mathbf{S}_{A,n} - \mathbf{S}_{C,n}' 
\vert^2 \right), 
\label{eqn:Scal}
\end{equation}
where the two-dimensional vectors, $\mathbf{S}_{A,n}$ and $\mathbf{S}_{C,n}'$, 
are the light rays in the initial (A) and recovered (C) source bundles. 

\begin{figure*}
\centering
\includegraphics[width=6.5in]{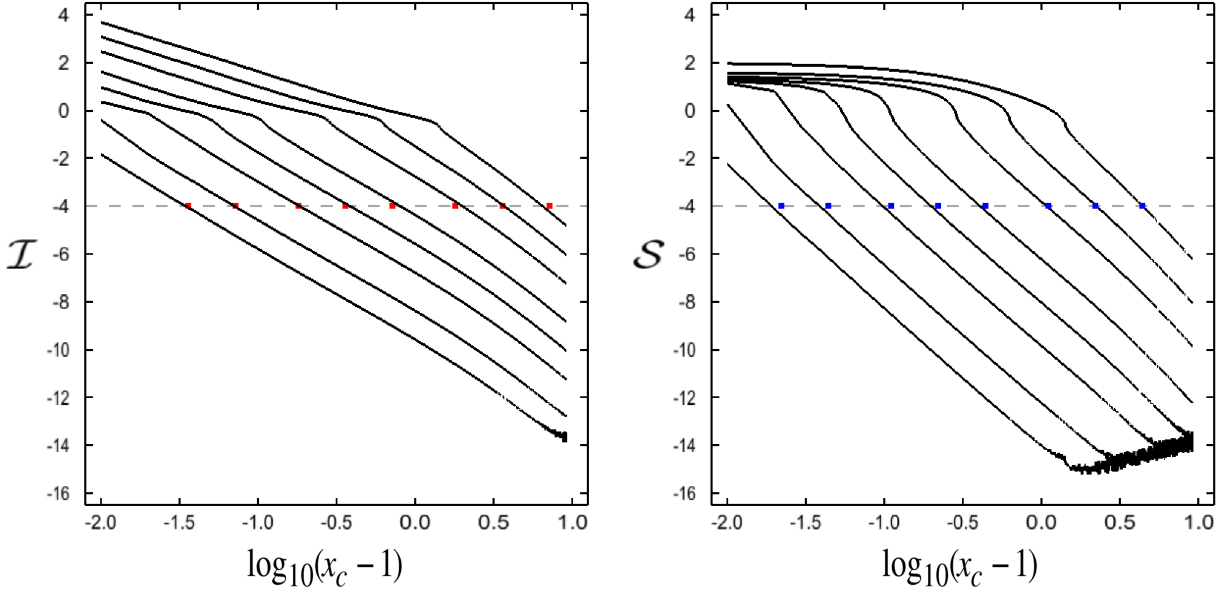}
\caption{${\cal I}$ (left) and ${\cal S}$ (right), as defined in 
equations (\ref{eqn:Ical}) and (\ref{eqn:Scal}), plotted as
a function of $\log_{10} (x_c - 1)$. Individual
lines are for different source radii (top to bottom) $\Delta y = 2.0, 1.0,
0.5,0.2,0.1, 0.05,0.02$ and $0.01$.  The horizontal dashed line in each
panel is at ${\cal I} = {\cal S} = -4$, taken as the limit above which
shape distortion is significant. The dots represent the empirical
fits (left) $x_{\rm hi} \gtrsim 1 + 3.6 \Delta y$ for ${\cal I}$ and
(right) $x_{\rm lo} \gtrsim 1 + 2.2 \Delta y$ for ${\cal S}$.
The noise in the lowest lines is due to 
the use of single precision floating point numbers. }
\label{fig:variation}
\end{figure*}

We calculate ${\cal I}$ and ${\cal S}$ for source bundles with radii 
$\Delta y$ = $0.01, 0.02, 0.05, 0.1, 0.2, 0.5, 1$ and 2.  
In Fig. \ref{fig:variation}, we plot ${\cal I}$ and ${\cal S}$ as 
functions of the image bundle impact parameter, $x_c$. 
We select ${\cal I} \geq -4$ and ${\cal S} \geq -4$ as indicative 
that the distortion is significant; this was confirmed by eye using plots 
similar to Fig. \ref{fig:faceplot}.  For higher values of ${\cal I}$ 
and ${\cal S}$, there were clear differences between sources 
A and C, and images B and D.
We find
\begin{equation}
x_c \gtrsim 1 + 2.2 \Delta y,
\label{eqn:thetalo1}
\end{equation}
for comparison of source bundle shapes A and C (${\cal S} \geq -4$) and
\begin{equation}
x_c \gtrsim 1 + 3.6 \Delta y,
\label{eqn:thetahi2}
\end{equation}
for comparison of image bundle shapes B and D (${\cal I} \geq -4$).

Next, we consider a more quantitative approach, based on comparing 
bundle ellipticities, in a manner comparable to the standard analysis 
for examining weak lensing-induced shear. 
Defining quadrupole terms 
\begin{equation}
Q_{ij} = \int \left(x_i - \bar{x}_i)(x_j - \bar{x}_j\right) {\rm d}^2 x
\end{equation}
relative to the bundle centroid
\begin{equation}
\bar{x}_i = \sum_{m=1}^{N_{\rm ray}} x_{i,m}
\end{equation}
we consider a complex ellipticity (e.g. Schneider 2005) of the form:
\begin{equation}
\chi \equiv \frac{Q_{11} - Q_{22} + 2 i Q_{12}}{Q_{11} + Q_{22}},
\end{equation}
which has norm:
\begin{equation}
\vert \chi \vert = \frac{\sqrt{(Q_{11}-Q_{22})^2 + 4 Q_{12}^2}}{Q_{11} + Q_{22}}.
\end{equation}
We define error terms:
\begin{eqnarray}
E_{AC} &=& \frac{\vert \chi \vert_A - \vert \chi \vert_C}{\vert \chi \vert_A}
\,\,\mbox{and}\\
E_{BD} &=& \frac{\vert \chi \vert_B - \vert \chi \vert_D}{\vert \chi \vert_B}
\label{eqn:eac}
\end{eqnarray}
for comparison between ellipticities, and determine the image plane
impact parameter, $x_c$, at which a source with radius, $\Delta y$, first 
exceeds $E = 1\%$, $5\%$ and $10\%$. 

Since a circular source has $\vert \chi \vert = 0$, we use 
elliptical sources with axis ratios $b/a = 0.8$, $0.9$ and $0.99$. 
We consider the two cases where the semi-major axis 
is aligned tangentially or radially to the Einstein radius, computing
these limits and also the average (based on original data values) 
of these two orientations.   

\begin{figure}
\begin{centering}
\includegraphics[width=3.4in]{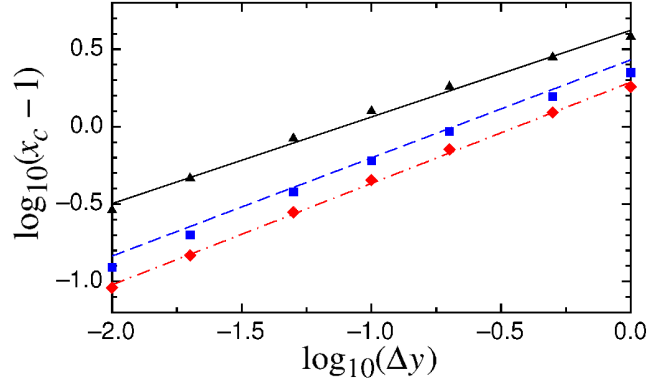}
\caption{Minimum impact parameter, $x_c$, at which the error, $E_{AC}$,
is first below $5\%$ as a function of the source radius, $\Delta y$, for 
elliptical source shapes with axis ratios $b/a = 0.8$ (red, dot-dashed line), 
$0.9$ (blue, dashed line) and $0.99$ (black, solid line). Lines are the 
least-squares fits in log-log space to the functional form
$x_c = 1 + \epsilon \Delta y^n$.  \label{fig:logger}} 
\end{centering}
\end{figure}

\begin{figure}
\begin{centering}
\includegraphics[width=3.4in]{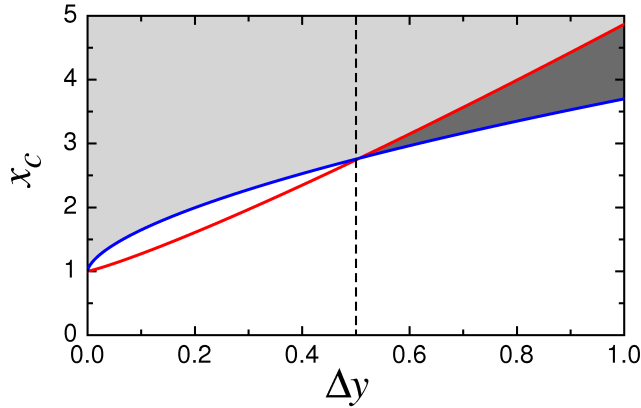}
\caption{The boundaries of the flexion zone (blue line) and the
shear zone (red line), as defined by equations (\ref{eqn:thetalo}) 
and (\ref{eqn:thetahi}) respectively.
The white region is the true strong lensing regime, while the light grey-shaded
region represents the preferred zone for both weak lensing shear and flexion
analysis.  The dark grey-shaded region, $\Delta y \gtrsim 0.5$ 
and hence source sizes comparable to the Einstein radius, means the 
stronger tidal fields ``flex'' rather than ``stretch'' images.
\label{fig:crossover}}
\end{centering}
\end{figure}

Plotting results as $\log_{10} (x_c - 1)$ versus $\log_{10} \Delta y$, 
see Fig.~\ref{fig:logger}, we see a 
relationship that is highly suggestive of a functional form:
\begin{equation}
x_c = 1 + \epsilon \Delta y^n.
\label{eqn:funcform}
\end{equation} 
We perform a least-squares fit in log-log space to obtain the 
parameters $\epsilon$ and $n$. Results of these fits are presented 
in Table \ref{tbl:lsf1} -- in all cases, the calculated Pearson
coefficient is $r > 0.994$, indicative that equation (\ref{eqn:funcform}) 
is an appropriate functional form.  There is variation in the
fitted parameters based on the chosen source axis ratio.  This is not
suprising, as the tidal gravitational field across the resultant image 
depends on the relative separations and orientation of individual image
rays from the lens (c.f. with discussion on orientation of image bundles
in section 3.4).   As a best estimate, we average over the three chosen
axis ratios for $E_{AC} = 5\%$, to obtain the second-to-last row in 
Table \ref{tbl:lsf1}.

\begin{table}
\caption{Least-squares fitting parameters for the functional
form $x_c = 1 + \epsilon \Delta y^n$ based on relative errors in source 
A and C ellipticities. Fits are made for sources with axis ratios $b/a = 0.8$,
0.9 and 0.99 with the semi-major axis aligned either tangentially or 
radially to the Einstein radius.  Fits were performed in log-log space, but
averaging is performed with original data values. 
In all cases, the Pearson coefficient is
$r \geq 0.994$. The inferred $\epsilon$ and $n$ values based on the 
limit ${\cal S} \geq -4$ are shown for comparison.}
\label{tbl:lsf1}
\centering
\begin{tabular}{ccccccccccc}
\hline
&& \multicolumn{2}{c}{Tangential} & \multicolumn{2}{c}{Radial} 
& \multicolumn{2}{c}{Average} \\
$b/a$ & $\mathbf{E_{AC}}$ & $\mathbf{\epsilon}$ & $\mathbf{n}$ 
& $\mathbf{\epsilon}$ & $\mathbf{n}$  
& $\mathbf{\epsilon}$ & $\mathbf{n}$ \\
\hline
\multirow{3}{*}{0.8} 
&$1\%$ & 3.58 & 0.65 & 2.14 & 0.56 & 2.86 & 0.61 \\
&$5\%$  & 2.27 & 0.67 & 1.60 & 0.63 & 1.94 & 0.65 \\
&$10\%$ & 1.87 & 0.68 & 1.35 & 0.65 & 1.61 & 0.67\\
\hline
\multirow{3}{*}{0.9}&1\% & 4.05 & 0.61 & 2.95 & 0.55 & 3.50 & 0.59 \\
&5\% & 2.69 & 0.65 & 2.16 & 0.62 & 2.43 & 0.63 \\
&10\% & 2.25 & 0.66 & 1.82 & 0.64 & 2.04 & 0.65 \\
\hline
\multirow{3}{*}{0.99}
&$1\%$ & 5.67 & 0.50 & 5.08 & 0.48 & 5.38 & 0.49 \\
&$5\%$ & 4.32 & 0.56 & 4.08 & 0.59 & 4.20 & 0.56 \\
&$10\%$ & 3.77 & 0.58 & 3.50 & 0.58 & 3.63 & 0.58 \\
\hline
Average& $5\%$ &&&&&2.70 & 0.62 \\
&${\cal S}$ & && &&  2.2 & 1.0  \\
\hline
\end{tabular}
\end{table}

\begin{table}
\caption{Least-squares fitting parameters for the functional
form $x = 1 + \epsilon \Delta y^n$ based on relative errors in image B 
and D ellipticities. In all cases, the Pearson coefficient 
is $r \geq 0.995$. The inferred
$\epsilon$ and $n$ values based on the limit ${\cal I} \geq -4$ are 
shown for comparison.}
\label{tbl:lsf2}
\centering
\begin{tabular}{ccc}
\hline
$\mathbf{E_{BD}}$ & $\mathbf{\epsilon}$ & $\mathbf{n}$ \\
\hline
1\% & 3.87 & 1.15 \\
5\% & 1.76 & 1.17 \\
10\% & 1.20 & 1.15 \\
${\cal I}$ & 3.6 & 1.0  \\
\hline
\end{tabular}
\end{table}

We propose the following interpretation: if we see an image that looks 
like B, 
and we use the flexion formalism to determine what the source would look 
like, we would be wrong (error of $\gtrsim 5\%$ in source ellipticity) unless:
\begin{equation}
x_c \gtrsim 1 + 2.7 \Delta y^{0.62}.
\label{eqn:thetalo}
\end{equation}
We refer to the boundary defined by this expression as the start of the
``flexion zone'' (blue line and both grey-shaded regions in Fig.~\ref{fig:crossover}).
For image impact parameters closer to the lens than this limit (white
region in Fig.~\ref{fig:crossover}), the second-order Taylor series 
approximation given by equation (\ref{eqn:Dijk}) is not sufficiently
accurate when applied to an extended source, and we are in the true
strong lensing regime.  The radius of the flexion zone boundary increases for 
larger source sizes, $\Delta y$, relative to the Einstein radius, which is 
expected as there will be greater tidal field variations across an 
image/source bundle.

We perform a similar analysis for
the relative ellipticity error between images B (RBM) and D (shear-only), 
although we now revert to using a circular source only.  Parameters are
presented in Table \ref{tbl:lsf2}.  We find that for
\begin{equation}
x_c \gtrsim 1 + 3.87 \Delta y^{1.15},
\label{eqn:thetahi}
\end{equation}
the lensed image shape and a shear-only intereptation are essentially
the same (error $\lesssim 1\%$ in ellipticity) implying the first-order 
Taylor expansion is adequate.  We refer to the boundary defined by
this expression as the start of the ``shear zone'',  where image 
shapes are essentially indistinguishable from ellipses, and a 
traditional weak-lensing (i.e. shear and convergence only) analysis 
is satisfactory.  However, our result does not preclude use of the 
flexion formalism at larger impact parameters, as a potentially 
measurable non-zero flexion remains. 

There is a crossover between the boundaries when $\Delta y \sim 0.5$, 
indicated by the dashed line in Fig.~\ref{fig:crossover}.  In some sense, 
the light grey-shaded region is the preferred region for weak lensing shear
and flexion analysis,
as it corresponds to both shear-only ellipticity errors $<1\%$ and 
flexion-recovered source ellipticity errors $<5\%$.  
The dark grey-shaded region, $\Delta y \gtrsim 0.5$ and hence source sizes 
comparable to the Einstein radius, means the stronger tidal fields ``flex'' 
rather than ``stretch'' images.
The effect of
this is demonstrated more clearly in Fig.~12, when we consider
cosmologically-realistic scenarios.

\begin{figure}
\includegraphics[width=3.4in]{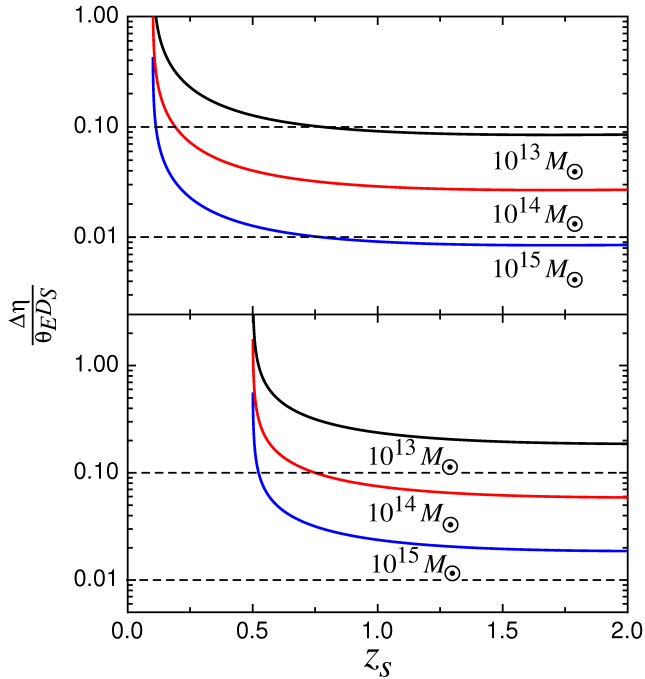}
\caption{The RBM-based constraint on bundle sizes for recovery of 
analytic flexion results applied to extended sources. 
For lens masses $M/M_{\odot} = 10^{13}, 10^{14}$ and $10^{15}$,
we plot the quantity $\frac{\Delta \eta}{\theta_{\rm E} D_{S}}$
for fixed $z_d = 0.1$ (top) and $z_d = 0.5$ (bottom), 
as a function of $z_s$, with $z_d < z_s \leq 2$. 
Results are for $\Delta \eta = 10$ kpc. The lower dashed line in 
each panel indicates $1 \%$ errors in recovered flexion, and 
the upper dashed line is for a $10 \%$ error.}
\label{fig:angsize}
\end{figure}

\section{Application}
\label{sct:discussion}
We approach our application of the results from the preceding section 
with the understanding that the Schwarzschild lens
is not an ideal description of the extended mass distribution of, for
example, a galaxy cluster lens.  Analytic predictions for first and second 
flexion do depend on the density profile of the lens model.  
Indeed, from Fig.~2 in Paper I, it can be seen that the Schwarzschild 
lens has zero $\kappa$ and $\vert {\cal F} \vert$, while these values 
are non-zero for the extended mass profiles (ie. SIS,  NFW and S\'{e}rsic 
profiles).  However, Birkhoff's theorem allows us to 
consider all truncated lens models to be Schwarzschild-like outside of
the truncation radius.  

For comparisons with observations, we need to convert our Einstein-radius
scaled results from Section \ref{sct:RBM} back to angular units on the sky.
With $\theta_i$ as the angular position in the lens plane and 
$\beta_i$ the angular position in the source plane, we have in the small
angle limit (which is appropriate for sources and lenses at cosmological
distances):
\begin{equation}
\theta_i = x_i \theta_{\rm E} \, \, \mbox{and} \, \, \beta_i = y_i \theta_{\rm E}
\end{equation}
where $\theta_{\rm E}$, the angular (point mass) Einstein radius, 
was defined in equation (\ref{eqn:apmerad}).

In the absence of lensing, a source at $\beta_i$ would be observed
at $\theta_i$, since from equation (\ref{eqn:angle}), $\beta_i = \theta_i$,
and the angular extent of the bundle satisfies $\Delta \beta = \Delta \theta$.
Substituting for $\Delta \beta = \Delta \eta/D_S$, equations 
(\ref{eqn:thetalo}) and (\ref{eqn:thetahi}) now become:
\begin{eqnarray}
x_{c} & \sim & 1 + \epsilon \left(\frac{\Delta \eta}{\theta_{\rm E} D_S}
\right)^{n},
\end{eqnarray}
or
\begin{eqnarray}
\theta_{c} & \sim & \theta_{\rm E} \left[ 1 + \epsilon 
\left(\frac{\Delta \eta}{ \theta_{\rm E} D_S}\right)^n
\right] \\
\label{eqn:scaler}
\end{eqnarray}

The RBM-limit on bundle sizes in angular units implies that the analytic 
flexion terms are most accurate for images with radii 
$\Delta \theta \lesssim 0.01 \theta_{\rm E}$.  We present results in 
Fig. \ref{fig:angsize}.
For lens masses $M/M_{\odot} = 10^{13}, 10^{14}$ and $10^{15}$,
we plot the quantity $\frac{\Delta \eta}{\theta_{\rm E} D_{S}}$
for fixed $z_d = 0.1$ (top) and $z_d = 0.5$ (bottom), 
as a function of $z_s$, with $z_d < z_s \leq 2$, and assume that 
$\Delta \theta \sim \Delta \beta$.  We choose $\Delta \eta = 10$ kpc, which is 
slightly smaller than the radius of a fiducial Milky Way-sized galaxy ($\Delta \eta
= 15$ kpc at $z_s = 0$), noting that the curves scale linearly with bundle radius.
We use the concordance cosmology, with total matter density, $\Omega_{\rm M,0} = 0.3$, 
dark energy density, $\Omega_{\Lambda,0} = 0.7$ and Hubble parameter, $H_0 =
100 h$ km s$^{-1}$ with h = 0.7.

The desired criteria (lower dashed line) at $0.01$ for a $1\%$ flexion error,
is only met for low lens redshifts and high mass lenses, 
typical of galaxy clusters, and for small source galaxy radii.
A more conservative limit at  $0.1$ (upper dashed line), results in a $10\%$ error
in recovered flexion values for extended sources -- see Fig.~\ref{fig:tau}.  While this 
may appear somewhat discouraging for flexion programs, we note that the required size 
criteria could be reached by considering isophotes of an image, corresponding to smaller
source sizes.  We do not discuss this further in the present work.

\begin{figure}
\begin{centering}
\includegraphics[width=3.4in]{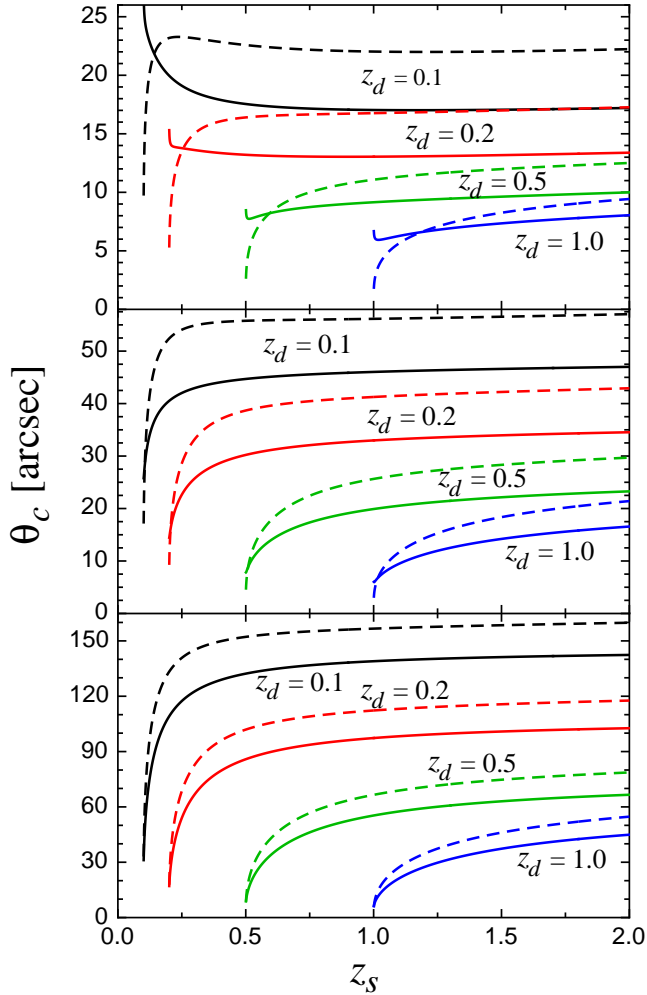} 
\end{centering}

\caption{The location of the inner boundaries of the flexion (dashed lines)
and shear zones (solid lines) for several typical scenarios.
Lens masses are $M = 10^{13} M_\odot$ (top row), $10^{14} M_\odot$ (middle row),
$10^{15} M_\odot$ (bottom row).  Lens redshifts in each panel are 
(from top to bottom) $z_d = 0.1$ (black), $0.2$ (red), $0.5$ (green) and $1.0$
(blue).  Source radius is $\Delta \eta = 10$ kpc. In most cases presented here, 
the shear zone commences closer to the lens than the flexion zone.}
\label{fig:3plot}
\end{figure}

\begin{figure}
\begin{centering}
\includegraphics[width=3.3in]{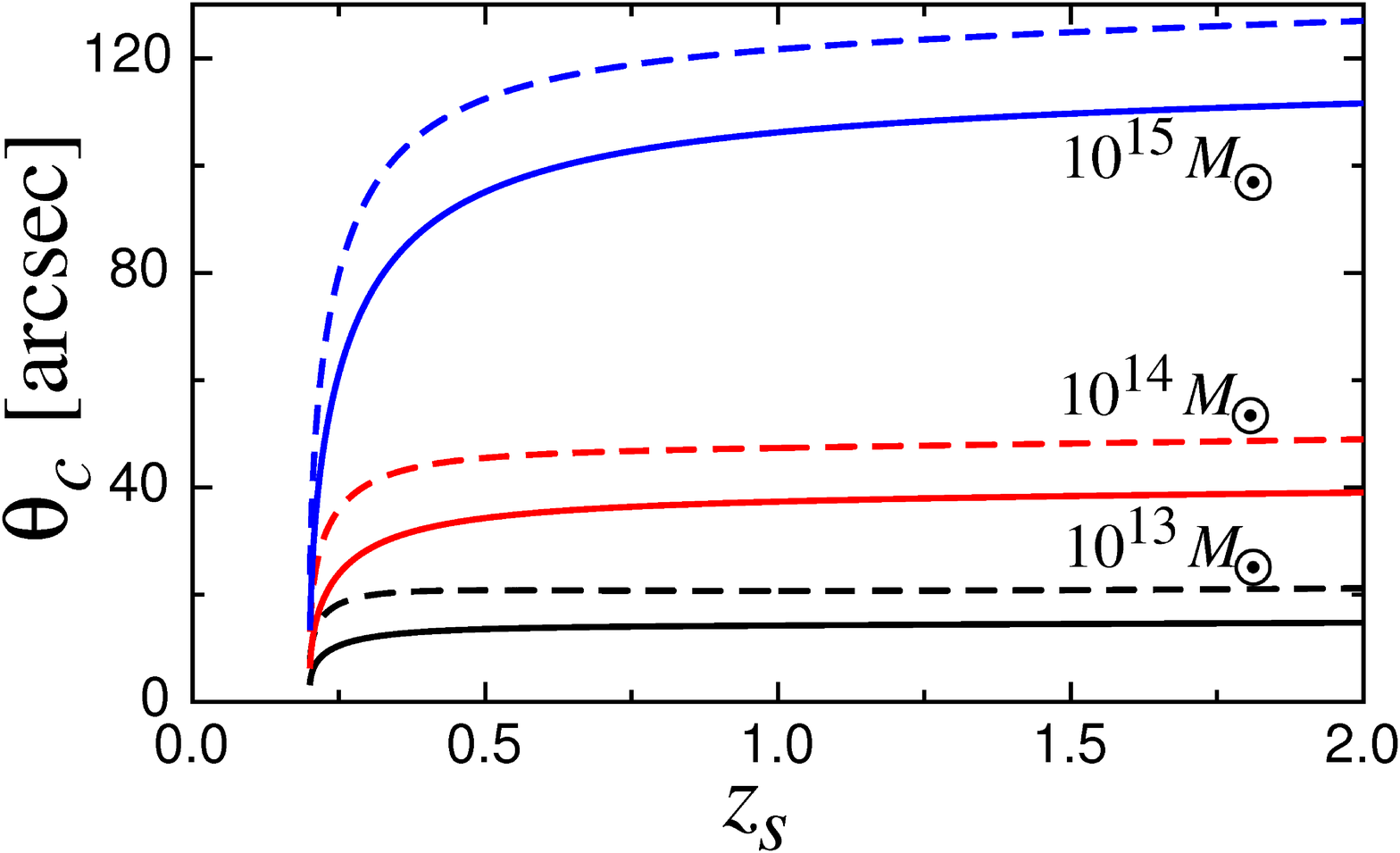} 
\end{centering}
\caption{Dependence of the flexion zone (arcsec) on the source radius, 
$\Delta \eta$, with lens redshift $z_{d} = 0.2$.  For each lens mass
the two lines are the inner boundaries of the flexion zone 
for $\Delta \eta = 20$ kpc (dashed lines) and $5$ kpc (solid lines).
Lens masses are $10^{13} M_\odot$ (bottom, black), $10^{14} M_\odot$ (middle, red)
and $10^{15} M_\odot$ (top, blue).}
\label{fig:rplot}
\end{figure}

In Fig.~\ref{fig:3plot}, 
we plot the location of the inner boundaries of the flexion zone (dashed
lines) and the shear zone (solid lines), for several typical scenarios:
lens masses $10^{13} M_{\odot}$ (top panel), $10^{14} M_{\odot}$ (middle panel) 
and $10^{15} M_{\odot}$ (bottom panel). In each panel (from top to bottom), 
the lens redshifts are $z_{d} = 0.1$ (black), $0.2$ (red), 
$0.5$ (green) and $1.0$ (blue), the source radius
is $\Delta \eta = 10$ kpc, and $z_{d} < z_s \leq 2$.  
Note the crossover between these regions
that occurs for $M = 10^{13} M_{\odot}$, and in most cases presented
here, the shear zone actually starts closer to the lens than the flexion zone.
Fig.~\ref{fig:rplot} shows the effect of changing the source
radius, $z_d = 0.2$, with $\Delta \eta = 20$ kpc (dashed lines) 
and $5$ kpc (solid lines).  
Lens masses used were $10^{13} M_{\odot}$ (black), $10^{14} M_\odot$ (red)
and $10^{15} M_{\odot}$ (blue).

\section{Concluding Remarks}
\label{sct:conclusion}
While weak lensing with shear is now well-established both theoretically and
observationally in the galaxy-galaxy lensing and cosmic shear cases, 
weak lensing via flexion is still in its infancy.

In Paper I, we considered analytic solutions for circularly-symmetric lens
models.  
In this paper, we have demonstrated how the ray-bundle method can be used
to recover the analytic second flexion results for the Schwarzschild
lens model to high accuarcy, and is consistent with the recovery 
of zero first flexion.   Indeed, we recover the Schwarzschild model
second flexion solutions with errors no worse than $1\%$ if bundle 
radii are $\Delta \theta \lesssim 0.01 \theta_{\rm E}$.  
In such circumstances, the second-order Taylor series expansion used 
by Bacon et al. (2006) is appropriate for extended sources.

Furthermore, we have identified the existence of a flexion zone in the 
image plane, which can be considered an optimal region for applying the 
analytic flexion formalism to extended sources.  

The ray-bundle method now provides us with a valuable numerical 
tool for studying 
flexion due to complex lens distributions, such as asymmetric lens 
models or cosmological structures, where no such analytic solutions exist.

\section{Acknowledgements}
This research was supported 
under Australian Research Council's Discovery Projects funding 
scheme (project number DP0665574).  PL is supported by the Alexander 
von Humboldt Foundation. Three-dimensional visualisation 
was conducted with the S2PLOT progamming library (Barnes et al. 2006).
We are grateful to Nick Bate for helpful comments on an earlier version
of this paper, and to the referee for insightful suggestions. 


\bsp

\label{lastpage}


\begin{thebibliography}{99}
\bibitem[\protect\citeauthoryear{Abate et al.}{2009}]{abate09}
Abate, A., Wittman, D., Margoniner, V. E., Bridle, S. L., Gee, P., Tyson, J. A., Dell'Antonio, I. P., 2009, ApJ, 702, 603

\bibitem[\protect\citeauthoryear{Bacon, Refregier \& Ellis}{2000}]{bacon00}
Bacon, D. J., Refregier, A. R., Ellis, R. S., 2000, MNRAS, 318, 625

\bibitem[\protect\citeauthoryear{Bacon et al.}{2003}]{bacon03}
Bacon, D. J., Massey, R. J., Refregier, A. R., Ellis, R. S., 2003, MNRAS, 344, 673

\bibitem[\protect\citeauthoryear{Bacon et al.}{2006}]{bacon06}
Bacon, D.J., Goldberg, D.M., Rowe, B.T.P., Taylor, A.N., 2006, MNRAS, 365, 414

\bibitem[\protect\citeauthoryear{Bacon, Amara \& Read}{2010}]{bacon09}
Bacon, D.J., Amara, A., Read, J.I., 2010, MNRAS, 409, 389

\bibitem[\protect\citeauthoryear{Bacon \& Sch\"{a}fer}{2009}]{bacon09b}
Bacon, D.J., Sch\"{a}fer, B.M., 2009, MNRAS, 396, 2167

\bibitem[\protect\citeauthoryear{Barnes et al.}{2006}]{barnes06}
Barnes, D.G., Fluke, C.J., Bourke, P.D., Parry, O.T., 2006, PASA, 13, 599

\bibitem[\protect\citeauthoryear{Brainerd, Blandford \& Smail}{1996}]{brainerd96}
Brainerd, T. G., Blandford, R. D., Smail, I., 1996, ApJ, 466, 623

\bibitem[\protect\citeauthoryear{Brown et al.}{2003}]{brown03}
Brown, M. L., Taylor, A. N., Bacon, D. J., Gray, M. E., Dye, S., Meisenheimer, K., Wolf, C., 2003, MNRAS, 341, 100

\bibitem[\protect\citeauthoryear{Dyer \& Roeder}{1974}]{dyer74}
Dyer, C.C., Roeder, R.C., 1974, ApJ, 189, 167

\bibitem[\protect\citeauthoryear{Fischer et al.}{2000}]{fischer00}
Fischer, P. et al., 2000, AJ, 120, 1198

\bibitem[\protect\citeauthoryear{Fluke et al.}{1999}]{fluke99}
Fluke, C.J., Webster, R.L., Mortlock, D.J., 1999, MNRAS, 306, 567

\bibitem[\protect\citeauthoryear{Fluke et al.}{2002}]{fluke02}
Fluke, C.J., Webster, R.L., Mortlock, D.J., 2002, MNRAS, 331, 180

\bibitem[\protect\citeauthoryear{Gavazzi \& Soucail}{2007}]{gavazzi07}
Gavazzi, R., Soucail, G., 2007, A\&A, 462, 459

\bibitem[\protect\citeauthoryear{Goldberg \& Bacon}{2005}]{goldberg05}
Goldberg, D. M., Bacon, D. J., 2005, ApJ, 619, 741

\bibitem[\protect\citeauthoryear{Goldberg \& Natarajan}{2002}]{goldberg02}
Goldberg, D. M., Natarajan, P., 2002, ApJ, 564, 65

\bibitem[\protect\citeauthoryear{Goldberg \& Leonard}{2007}]{goldberg07}
Goldberg, D.M., Leonard, A., 2007, ApJ, 660, 1003

\bibitem[\protect\citeauthoryear{Gray et al.}{2002}]{gray02}
Gray, M. E., Taylor, A. N., Meisenheimer, K., Dye, S., Wolf, C., Thommes, E., 2002, ApJ, 568, 141

\bibitem[\protect\citeauthoryear{Guzik \& Seljak}{2002}]{guzik02}
Guzik, J., Seljak, U., 2002, MNRAS, 335, 311

\bibitem[\protect\citeauthoryear{Harper}{1991}]{harper91}
Harper, J.F.P., 1991, PhD Thesis, University of Toronto

\bibitem[\protect\citeauthoryear{Hawken \& Bridle}{2009}]{hawken09}
Hawken, A.J., Bridle, S.L., 2009, MNRAS, 400, 1132

\bibitem[\protect\citeauthoryear{Heavens, Kitching \& Taylor}{2006}]{heavens06}
Heavens, A. F., Kitching, T. D., Taylor, A. N., 2006, MNRAS, 373, 105

\bibitem[\protect\citeauthoryear{Heymans et al.}{2006}]{heymans06}
Heymans, C. et al., 2006, MNRAS, 371, L60

\bibitem[\protect\citeauthoryear{Hoekstra et al.}{2002}]{hoekstra02}
Hoekstra, H., Yee, H. K. C., Gladders, M. D., Barrientos, L. F., Hall, P. B., Infante, L., 2002, ApJ, 572, 55

\bibitem[\protect\citeauthoryear{Hoekstra, Yee \& Gladders}{2004}]{hoekstra04}
Hoekstra, H., Yee, H. K. C., Gladders, M. D., 2004, ApJ, 606, 67

\bibitem[\protect\citeauthoryear{Hoekstra \& Jain}{2008}]{hoekstra08}
Hoekstra, H., Jain, B., 2008, Ann. Rev. of Nuclear and Particle Science, 58, 99

\bibitem[\protect\citeauthoryear{Hudson et al.}{1998}]{hudson98}
Hudson, M. J., Gwyn, S. D. J., Dahle, H., Kaiser, N., 1998, ApJ, 503, 531

\bibitem[\protect\citeauthoryear{Irwin \& Shmakova}{2005}]{irwin05}
Irwin, J., Shmakova, M., 2005, New Astron. Rev., 49, 53

\bibitem[\protect\citeauthoryear{Irwin \& Shmakova}{2006}]{irwin06}
Irwin, J., Shmakova, M., 2006, ApJ, 645, 17

\bibitem[\protect\citeauthoryear{Irwin, Shmakova \& Anderson}{2007}]{irwin07}
Irwin, J., Shmakova, M., Anderson, J., 2007, ApJ, 671, 1182

\bibitem[\protect\citeauthoryear{Johnston et al.}{2007}]{johnston07}
Johnston, D. E., Sheldon, E. S., Tasitsiomi, A., Frieman, J. A., Wechsler, R. H., McKay, T. A., 2007, ApJ, 656, 27

\bibitem[\protect\citeauthoryear{Kaiser}{1995}]{kaiser95}
Kaiser, N., 1995, ApJ, 439, L1

\bibitem[\protect\citeauthoryear{Kayser}{1986}]{kayser86}
Kayser, R., Refsdal, S., Stabell, R., 1986, A\&A, 166, 36


\bibitem[\protect\citeauthoryear{Kitching et al.}{2007}]{kitching07}
Kitching, T. D., Heavens, A. F., Taylor, A. N., Brown, M. L., Meisenheimer, K., Wolf, C., Gray, M. E., Bacon, D. J., 2007, MNRAS, 376, 771

\bibitem[\protect\citeauthoryear{Lasky \& Fluke}{2009}]{lasky09}
Lasky, P. D., Fluke, C. J., 2009,  MNRAS, 396, 2257 (Paper I)

\bibitem[\protect\citeauthoryear{Leonard et al.}{2007}]{leonard07}
Leonard, A., Goldberg, D. M., Haaga, J. L., Massey, R., 2007, ApJ, 666, 51

\bibitem[\protect\citeauthoryear{Leonard \& King}{2010}]{leonard10}
Leonard, A. King, L.J., 2010, MNRAS, 405, 1854 

\bibitem[\protect\citeauthoryear{Leonard, King \& Wilkins}{2009}]{leonard09}
Leonard, A., King, L. J., Wilkins, S. M., 2009, MNRAS, 395, 1438

\bibitem[\protect\citeauthoryear{Massey et al.}{2007}]{massey07}
Massey, R., Rowe, B., Refregier, A., Bacon, D.J., Berg\'{e}, J., 2007, MNRAS, 380, 229

\bibitem[\protect\citeauthoryear{Mandelbaum et al.}{2006}]{mandelbaum06}
Mandelbaum, R., Hirata, C. M., Broderick, T., Seljak, U., Brinkmann, J., 2006, MNRAS, 370, 1008

\bibitem[\protect\citeauthoryear{Mandelbaum, Seljak \& Hirata}{2008}]{mandelbaum08}
Mandelbaum, R., Seljak, U., Hirata, C. M., 2008, JCAP, 8, 6

\bibitem[\protect\citeauthoryear{Okabe et al.}{2010}]{okabe09}
Okabe, N., Takada, M., Umetsu, K., Futamase, T., Smith, G. P., 2010, 
PASJ, 62, 811

\bibitem[\protect\citeauthoryear{Okura et al.}{2007}]{okura07}
Okura, Y., Umetsu, K., Futamase, T., 2007, ApJ, 660, 995 

\bibitem[\protect\citeauthoryear{Okura et al.}{2008}]{okura08}
Okura, Y., Umetsu, K., Futamase, T., 2008, ApJ, 680, 1

\bibitem[\protect\citeauthoryear{Paczy\'{n}ski}{1986}]{paczynski86}
Paczy\'{n}ski, B., 1986, ApJ, 301, 503 

\bibitem[\protect\citeauthoryear{Parker et al.}{2007}]{parker07}
Parker, L. C., Hoekstra, H., Hudson, M. J., van Waerbeke, L., Mellier, Y., 2007 ApJ, 669, 21


\bibitem[\protect\citeauthoryear{Refregier, Rhodes \& Groth}{2002}]{refregier02}
Refregier, A., Rhodes, J., Groth, E. J., 2002, ApJ, 572, L131

\bibitem[\protect\citeauthoryear{Rhodes, Refregier \& Groth}{2001}]{rhodes01}
Rhodes, J., Refregier, A., Groth, E. J., 2001, ApJ, 552, L85


\bibitem[\protect\citeauthoryear{Sachs}{1961}]{sachs61}
Sachs, R.K., 1961, Proc. Roy. Soc. London, A, 264, 309

\bibitem[\protect\citeauthoryear{Schneider}{2005}]{schneider05}
Schneider, P., 2005, in Jetzer, P., North, P., eds, Gravitational Lensing: Strong, Weak and Micro Weak Gravitational Lensing, Springer-Verlag, Berlin

\bibitem[\protect\citeauthoryear{Schneider \& Er}{2008}]{schneider08}
Schneider, P., Er, X., 2008, A\&A, 485, 363

\bibitem[\protect\citeauthoryear{Schneider \& Weiss}{1986}]{schneider86}
Schneider, P., Weiss, A., 1986, A\&A, 164, 237

\bibitem[\protect\citeauthoryear{Schneider \& Weiss}{1987}]{schneider87}
Schneider, P., Weiss, A., 1987, A\&A, 171, 49

\bibitem[\protect\citeauthoryear{Shapiro et al.}{2010}]{shapiro10}
Shapiro, C., Bacon, D., Hendry, M., Hoyle, B., 2010, MNRAS, 404, 858

\bibitem[\protect\citeauthoryear{Sheldon et al.}{2004}]{sheldon04}
Sheldon, E. S. et al., 2004, AJ, 127, 2544

\bibitem[\protect\citeauthoryear{Smail et al.}{1997}]{smail97}
Smail, I., Ellis, R. S., Dressler, A., Couch, W. J., Oemler, A., Sharples, R. M., Butcher, H., 1997, ApJ, 479, 70

\bibitem[\protect\citeauthoryear{Smith et al.}{2001}]{smith01}
Smith, D. R., Bernstein, G. M., Fischer, P., Jarvis, M., 2001, ApJ, 551, 643

\bibitem[\protect\citeauthoryear{Taylor et al.}{2004}]{taylor04}
Taylor, A. N., Bacon, D. J., Gray, M. E., Wolf, C., Meisenheimer, K., Dye, S., Borch, A., Kleinheinrich, M., Kovacs, Z., Wisotzki, L., 2004, MNRAS, 353, 1176

\bibitem[\protect\citeauthoryear{Tyson et al.}{1984}]{tyson84}
Tyson, J. A., Valdes, F., Jarvis, J. F., Mills, A. P., ApJ, 1984, 281, L59

\bibitem[\protect\citeauthoryear{Valdes et al.}{1983}]{valdes83}
Valdes, F., Jarvis, J. F., Tyson, J. A., ApJ, 1983, 271, 431 

\bibitem[\protect\citeauthoryear{Wambsganss}{1990}]{wambsganss90}
Wambsganss, J., 1990, PhD Thesis, MPA Report 550

\bibitem[\protect\citeauthoryear{Wambsganss}{1999}]{wambsganss99}
Wambsganss, J., 1999, J. Comput. Appl. Math., 109, 353 


\bibitem[\protect\citeauthoryear{Wittman et al.}{2000}]{wittman00}
Wittman, D., Tyson, J. A., Kirkman, D., Dell'Antonio, I., Bernstein, G., 2000, 
Nat, 405, 143

\bibitem[\protect\citeauthoryear{Wittman et al.}{2001}]{wittman01}
Wittman, D., Tyson, J. A., Margoniner, V. E., Cohen, J. B., Dell'Antonio, I. P., 2001, ApJ, 557, L89


\end{thebibliography}
\end{document}